\documentclass[bibyear]{aa}

\usepackage{graphicx}
\usepackage{txfonts}
\usepackage[utf8]{inputenc}
\usepackage{amsfonts}
\usepackage{amsmath}
\usepackage{bm}
\usepackage{xcolor}
\usepackage{color}
\usepackage{amssymb}
\usepackage{natbib}

\usepackage{ulem}

\newcommand{\de}{\partial}
\newcommand{\bnabla}{{\bm \nabla}}
\newcommand{\bx}{{\bm x}}
\newcommand{\br}{{\bm r}}
\newcommand{\bv}{{\bm v}}
\newcommand{\bb}{{\bm B}}
\newcommand{\ba}{{\bm a}}
\newcommand{\bj}{{\bm j}}

\usepackage{orcidlink}

\begin{document}

\title{Turbulence and particle energization in twisted flux ropes in solar-wind conditions}

\author{O. Pezzi\orcidlink{0000-0002-7638-1706}\inst{1}, 
D. Trotta\orcidlink{0000-0002-0608-8897}\inst{2}, 
S. Benella\orcidlink{0000-0002-7102-5032}\inst{3}, 
L. Sorriso-Valvo\orcidlink{0000-0002-5981-7758}\inst{1,4},
F. Malara\orcidlink{0000-0002-5554-8765}\inst{5,6}, 
F. Pucci\orcidlink{0000-0002-5272-5404}\inst{1}, 
C. Meringolo\orcidlink{0000-0001-8694-3058}\inst{5}
W.H. Matthaeus\orcidlink{0000-0001-7224-6024}\inst{7}, \& 
S. Servidio\orcidlink{0000-0001-8184-2151}\inst{5,6}}

\authorrunning{O. Pezzi et al.}
\institute{Istituto per la Scienza e Tecnologia dei Plasmi (ISTP), Consiglio Nazionale delle Ricerche, Via Amendola 122/D, I-70126 Bari, Italy\\ \email{oreste.pezzi@istp.cnr.it}
\and  
The Blackett Laboratory, Department of Physics, Imperial College London, London SW7 2AZ, UK
\and
Istituto di Astrofisica e Planetologia Spaziali, Istituto Nazionale di Astrofisica, via del Fosso del Cavaliere 100, I-00133, Rome, Italy
\and
Space and Plasma Physics, School of Electrical Engineering and Computer Science, KTH Royal Institute of Technology, Teknikringen 31, SE-11428 Stockholm, Sweden  
\and
Dipartimento di Fisica, Università della Calabria, Ponte P. Bucci, Cubo 31C, I-87036 Arcavacata di Rende (CS), Italy
\and
Istituto Nazionale di Astrofisica - INAF, Direzione Scientifica, Roma (Italy)
\and
Bartol Research Institute and Department of Physics and Astronomy, University of Delaware, Newark, DE 19716, USA
}

\date{\today}

\abstract
   {The mechanisms regulating the transport and energization of charged particles in space and astrophysical plasmas are still debated. Plasma turbulence is known to be a powerful particle accelerator. Large-scale structures, including flux ropes and plasmoids, may contribute to confine particles and lead to fast particle energization. These structures may also modify the properties of the turbulent nonlinear transfer across scales.}
   {We investigate how large-scale flux ropes are perturbed and, simultaneously, influence the nonlinear transfer of turbulent energy towards smaller scales. We then address how these structures affect particle transport and energization.
   %\ser{We investigate the interaction between large-scale magnetic flux ropes and turbulence in heliospheric conditions. We then address how these complex scenarios influence the process of particle energization.}
   }
   {We adopt magnetohydrodynamic simulations for perturbing a large-scale flux rope in solar-wind conditions and possibly triggering turbulence. Then, we employ test-particle methods to investigate particle transport and energization in the perturbed flux rope.}
   {The large-scale helical flux rope inhibits the turbulent cascade towards smaller scales, especially if the amplitude of the initial perturbations is not large ($\sim 5\%$). In this case, particle transport is inhibited inside the structure. Fast particle acceleration occurs in association with phases of trapped motion within the large-scale flux rope.}
   {}

\maketitle

\section{Introduction}
\label{sect:intro}

Charged particles energized to extremely high energies are ubiquitous in the Universe, as shown by direct observations in the heliosphere \citep{reames99particle, dresing202317} and inferred from remote observations of far astrophysical systems \citep{lazarian2012turbulence, amato2021particle, cristofari2021hunt}. Despite decades of research on the topic, starting 
from the seminal works by \citet{Fermi1949, Fermi1954}, 
the mechanisms responsible for particle acceleration in nearly-collisionless plasmas are still elusive.

Among different fundamental processes leading to efficient particle energization \citep{fisk2012particle,retino2022particle}, large-scale plasma turbulence has been central in different recent theoretical and numerical efforts \citep[e.g.,][and references therein]{lazarian20203D}. In particular, intermittency gives rise to inhomogeneous patches of coherent structures, such as vortices, current sheets, plasmoids and flux ropes, across a vast range of spatial scales \citep{matthaeus2015intermittency, marino2023scaling}. 
In the solar wind, these locally-generated plasma structures, whose origin is possibly associated with magnetic reconnection, are often complemented by structures of solar origin traveling in the heliosphere \citep[e.g.,][]{malandraki2019current}. As a matter of fact, flux-rope-like structures are routinely observed in the heliosphere at different distances from the Sun \citep[][and references therein]{hu2014structures, hu2018automated, khabarova2021currentsheets,reville2022flux}.

A first interesting aspect concerns the dynamical evolution of ``mesoscale'' structures observed at both large (fluid) and small (kinetic) scales \citep{viall2021mesoscale}. How mesoscale structures are affected by the turbulent background in which they travel and, at the same time, mediate the cascade of turbulence towards smaller scales is still a puzzle. Different studies, performed in the context of solar coronal loops, investigated the propagation of magnetohydrodynamics (MHD) waves and the onset of instabilities in magnetic flux tubes. \citet{emonet1998physics} explored the dynamics of twisted flux ropes in a stratified medium mimicking their emergence from the solar convection zone, while \citet{srivastava2010observation} reported the evidence of kink instability in the context of a solar flare. The propagation of kink and torsional Alfv\'en waves, relevant for coronal heating through, for example, resonant absorption or the generation of the Kelvin-Helmholtz instability at the boundary of coronal loops, has been extensively investigated \citep{terradas2008nonlinear,antolin2010role, antolin2014fine, magyar2015numerical, karampelas2017heating, howson2017energetics, antolin2017observational}. More recently, \citet{diazsuarez2021transition,diazsuarez2022transition} described the transition from linear phase-mixing to turbulence ---mediated by Kelvin-Helmholtz instability--- in a coronal loop perturbed by torsional Alfv\'en waves. \citet{diazsuarez2022transition} also examined the role of the twist in the magnetic field structure of the coronal loops, finding that the twist has a stabilizing effect on the Kelvin-Helmholtz instability, thus preventing the transition to the turbulent behavior.

A second challenging issue consists in understanding how these structures mediate the transport and energization of particles. In the solar corona, flux ropes ---often associated with solar flares--- are known to produce intense X-ray emission \citep{pinto2015soft,pinto2016thermal}, thus being potential sites of intense particle acceleration. A different perspective highlights the role of large-scale structures, of which flux ropes are a sub-ensemble, in trapping particles. Trapped particles may be quickly energized, as shown in numerous numerical and theoretical efforts that investigated the role of magnetic reconnection in generating islands and plasmoids whose merging or contraction leads to efficient particle energization \citep{drake2006electron, oka2010electron, kowal2011magnetohydrodynamic, kowal2012particle, leroux2015kinetic, leroux2018selfconsistent, leroux2019modeling,pezzi2021currentsheets}. The role of particle trapping has been also explored in the context of magnetic discontinuities \citep{malara2021charged} and switchbacks \citep{malara2023energetic}. More generally, plasma turbulence produces coherent structures possibly trapping particles and leading to fast particle acceleration \citep{Dmitruk_al_2003, DmitrukEA04, drake2006electron, servidio2016explosive, pecora2017ion, pisokas2018synergy, trotta2020fast, pezzi2022relativistic, lemoine2022first}. In this case, the acceleration process is quite complex and characterized by the interplay between stochastic acceleration due to turbulent MHD fluctuations and experienced by all the particles and, possibly, systematic energization associated with trapping and perceived by a relatively small number of particles \citep{ambrosiano1988test, pezzi2022relativistic}. These extensive works have been complemented by observational analyses studying particle acceleration in small-scale flux ropes \citep{khabarova2015smallscale, khabarova2016smallscale} as well as large-scale flux-tube structures \citep{pecora2021parker,mccomas2023parker}. In particular, \citet{pecora2021parker} confirmed that twisted flux tubes are a transport barrier for energetic particles which, as a result, are confined within or at the boundary of the flux tube itself (see also \citep{tooprakai2007temporary, krittinatham2009drift, tooprakai2016simulations}. In the context of particle acceleration at shocks, turbulent structures have been found to be responsible for additional energization due to their capability to trap particles and, more generally, influence their transport properties~\citep{Zank2015}. This complex picture about the interplay between shocks and turbulent structures has been emerging in recent theoretical~\citep{Zank2021}, numerical~\citep{Nakanotani2021, Trotta2022a} and observational~\citep{Kilpua2023sheath} studies.

This work furthers the present understanding of the two above discussions. In particular, we address (a) how plasma turbulence is influenced and perturbs a large-scale flux rope and (b) how the perturbed flux rope affects particle transport and energization, using a combination of MHD and test-particle simulations. The MHD approach is adopted to investigate the development of plasma turbulence in presence of a large-scale flux rope, generated as a Grad-Shafranov equilibrium with parameters in qualitative accordance with those observed in the solar wind \citep{hu2018automated}. Test-particle simulations are then employed to study particle transport and energization in the perturbed flux rope. 

By performing different runs in which the flux rope is perturbed with large-scale fluctuations, whose initial energy is varied within two orders of magnitude, we show that the turbulent cascade within the flux rope is generally inhibited. As the amplitude of initial perturbations increases, this effect vanishes as turbulence dominates on the effect of the large-scale magnetic structure. 
Then, we investigate how the turbulent flux rope influences particle transport and energization by performing test-particle simulations under a stationary assumption. Our main findings are that, in cases of small amplitude of perturbations, particle transport is inhibited within the structure which can entrap particles. While trapped, particles can experience an efficient acceleration due to intense electric fields.

The structure of the paper is the following. Section \ref{sect:nummod} describes the adopted numerical models and the setup of numerical simulations. Section \ref{sect:turb} discusses how turbulence affects the flux rope and what differences arise in the turbulent cascade inside or outside the structure. Then, Section \ref{sect:acc} analyzes how particle transport and acceleration is influenced by the presence of the flux rope. In Section \ref{sect:concl} we conclude by summarizing our findings.

\section{Numerical models and simulations' setup}
\label{sect:nummod}
The numerical method adopted in the current work combines a compressible MHD algorithm and a test-particle code. The MHD code is exploited to perturb the initial equilibrium which is characterized by a flux rope built through the Grad-Shafranov technique. Test-particle methods allow us to analyze particle transport and acceleration in the perturbed turbulent flux rope. We here provide details about the numerical models and simulations' setup.

\subsection{Compressible MHD solver}
\label{sect:nummod_MHD}

\begin{figure}[!htb]
 \centering
\begin{minipage}{0.35\textwidth}
\includegraphics[width=\textwidth]{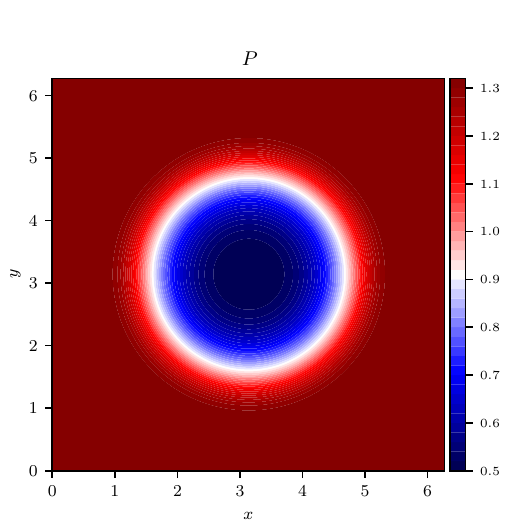}
\end{minipage}
\begin{minipage}{0.35\textwidth}
\includegraphics[width=\textwidth]{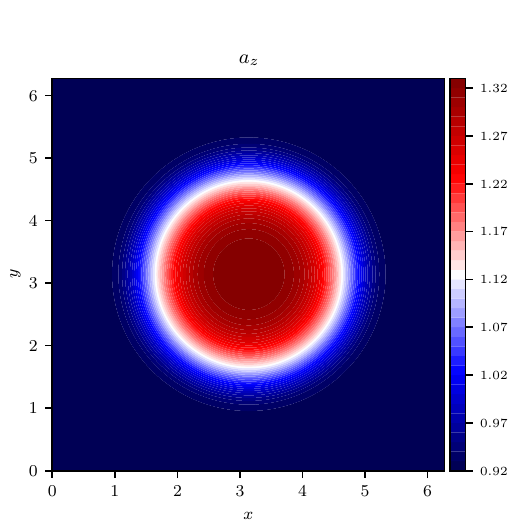}
\end{minipage}
\caption{Overview of the initial two-dimensional equilibrium, showing the maps of the kinetic pressure $P$ (top) and the out-of-plane potential vector $a_z$ (bottom).}
\label{fig:EQ_2Dmaps}
\end{figure}

The MHD code numerically integrates the three-dimensional (3D) equations of the compressible magnetohydrodynamics that, in normalized units, can be written as:
\begin{eqnarray}
&& \frac{\de \rho}{\de t} = - \bnabla \cdot \left( \rho {\bm v} \right) \label{eq:MHDrho} \\
&& \frac{\de \bv}{\de t} = - \left( \bv \cdot \bnabla \right) \bv + \frac{1}{\rho} \left[ \left(\bnabla \times \bb\right)\times\bb \right] - \frac{\beta_0}{2\rho}\bnabla\left(\rho T\right) \label{eq:MHDv} \\
&& \frac{\de \bb}{\de t} = \bnabla \times \left( \bv \times \bb  \right) \label{eq:MHDB} \\
&& \frac{\de T}{\de t} = - \left( \bv \cdot \bnabla \right) T - \left( \gamma -1\right) \left(\bnabla \cdot \bv \right) T  \label{eq:MHDT}
\end{eqnarray}
where $(\rho,\bv,T,\bb)(\bx,t)$ are respectively the magnetofluid density, velocity, temperature, and the magnetic field, while $\gamma=5/3$ is the adiabatic index. Equations (\ref{eq:MHDrho}--\ref{eq:MHDT}) are normalized as follows. Lengths, times, and velocities are respectively scaled to the energy-containing scale $L_A$, Alfv\'en crossing time $t_A=L_A/v_A$, and Alfv\'en speed $v_A = B_0/\sqrt{4\pi \rho_0}$, evaluated with the normalizing magnetic field $B_0$ and density $\rho_0$. The algorithm includes also the Hall and electron pressure terms in the induction equation for the magnetic field, turned off for this study. 

The normalized pressure has been written as $P=\beta_0 \rho T/2$ where $\beta_0$ is the kinetic to magnetic pressure ratio evaluated with the normalizing quantities $n_0$, $B_0$, and $T_0$, i.e., in cgs units $\beta_0=n_0 k_B T_0/(B_0^2/8\pi)$.
This ensures that normalized density and temperature are order $\sim 1$, and, in normalized units, $\beta=P/(B^2/2)=\beta_0 \cdot \left(\rho T/B^2\right)$. Hence, the effective plasma $\beta$ can actually change in the computational domain due to either non-homogeneous equilibrium quantities or possible perturbations. We here set $B_0=5 {\rm nT}$, $n_0=10 {\rm cm}^{-3}$, $L_A=5\times 10^3 {\rm km}$, and $T_0\simeq 1.4 \times 10^{5} K$, thus providing $v_A\simeq 35 {\rm km/s}$ and $\beta_0=2$. 

Equations (\ref{eq:MHDrho}--\ref{eq:MHDT}) are integrated prescribing a logarithmic regularization to the density $\rho\equiv e^{g}$ and solving the equivalent equation for $g$ with the purpose of better describing possible discontinuities and shocks. Moreover, we integrate the MHD equations for the magnetic potential $\ba$ in place of the magnetic field $\bb=\bnabla \times \ba$ to guarantee $\nabla \cdot \bb = 0$. 

MHD equations are integrated on a tri-periodic cube of size is $L_x=L_y=L_z=L=2 \pi L_A$, discretized with $N_x=N_y=N_z=N=512$ gridpoints in each direction. The algorithm adopts a pseudo-spectral method \citep{canuto2006spectral} based on Fast Fourier Transform (FFTW) routines  \citep{frigo1999fast} to compute the right-hand side of MHD equations: spatial derivatives are computed in the Fourier space, while products between variables are calculated in physical space. Then, the time evolution is performed through a second-order Runge-Kutta scheme. 
In order to ensure numerical stability, we adopt the standard dealiasing procedure at $k_{\rm alias}=2/3 k_{\rm max}$ ($k_{\rm max}=N/2$) and implement hyper-viscous terms ($\propto \nabla^4 $) in Eqs. (\ref{eq:MHDv}--\ref{eq:MHDT}), being the hyper-viscous coefficients $\nu\simeq 10^{-8}$. 
The code, dubbed as \texttt{COHMPA} (``COmpressible Hall Magnetohydrodynamics simulations for Plasma Astrophysics''), employs a domain-parallelization strategy based on the MPI paradigm. A 2.5D version of this algorithm has been already adopted in literature \citep{vasconez2015kinetic, perri2017numerical, pezzi2017revisiting}. 

\begin{figure}[!t]
\centering
\includegraphics[width=\columnwidth]{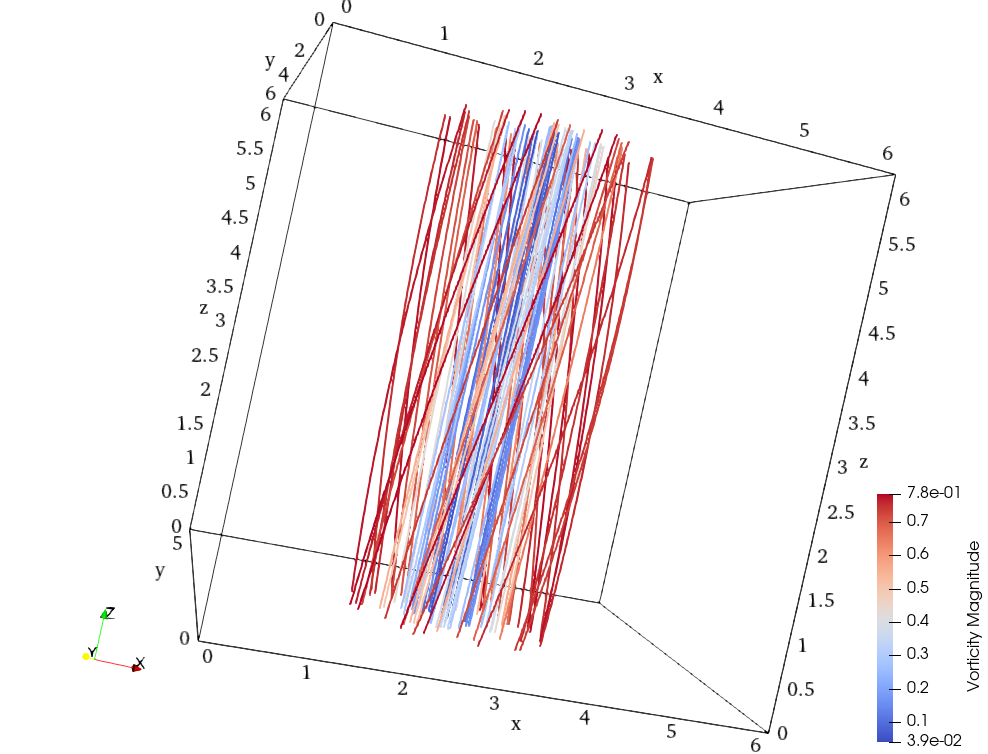}
\caption{Three-dimensional magnetic field lines of the equilibrium flux-rope structure. Lines are coloured with the magnitude of the magnetic field vorticity, i.e., the current density.}
\label{fig:equil3D}
\end{figure}

\begin{figure}[!t]
\centering
\includegraphics[width=0.9\columnwidth]{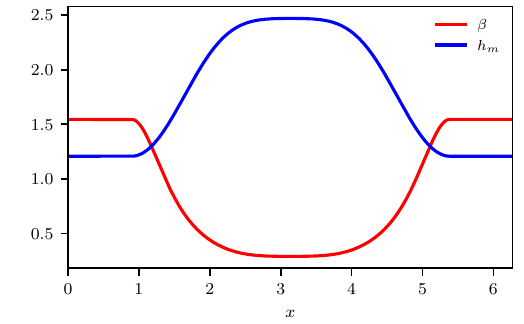}
\caption{One-dimensional profile of $\beta$ (red) and the density of magnetic helicity $h_m$ (blue) along $x$, evaluated at $y=L_y/2$.}
\label{fig:EQ_beta}
\end{figure}

MHD simulations are initialized by perturbing a two-dimensional Grad--Shafranov equilibrium which produces a flux rope. The details of the Grad-Shafranov technique are reported in App. \ref{App:GS}. The initial condition is characterized by a decrease of kinetic pressure $P$, and an increase of the out-of-plane magnetic vector potential $a_z$ inside the structure with respect to the surrounding environment, as shown in Fig. \ref{fig:EQ_2Dmaps}. The flux-rope width is $L_{\rm FR}\simeq 2 L_A$, while the scale associated with flux-rope gradients is $L_{\rm \Delta}\simeq L_A$. As described in App. \ref{App:GS}, the magnetic field associated with the magnetic vector potential $a_z$ has a dipolar structure in the plane, while its $z$-component increases in the flux rope with respect to the external region. Owing to the presence of a parallel component in the equilibrium magnetic field, magnetic field lines twist along the flux-rope axis. Such a behavior can be appreciated in Fig. \ref{fig:equil3D} which displays the magnetic field lines, colored with the intensity of the current density, $\bj=\nabla\times\bb$. 

This configuration is often observed in the solar wind, although the solar wind exhibits a large variability in terms of flux-rope parameters and overall configuration \citep{hu2018automated}. The plasma $\beta$ changes from the external value  $\sim 1.5$ to the internal value $\sim 0.3$ (red line in Fig. \ref{fig:EQ_beta}), while the local density of magnetic helicity $h_m=\ba\cdot\bb$ increases in the structure (blue line in Fig. \ref{fig:EQ_beta}). 

The initial equilibrium is perturbed with large-scale fluctuations of magnetic field $\delta{\bm  b}$ and bulk speed $\delta{\bm  v}$ such that $H_c=\langle \delta{\bm v} \cdot \delta{\bm b} \rangle \simeq 0$ and $H_m=\langle \delta{\bm a} \cdot \delta{\bm b} \rangle \simeq 0$, being $\delta {\bm b}=\bnabla \times \delta {\bm a}$. These fluctuations are fully 3D polarized, $\delta b_x \sim \delta b_y \sim \delta b_z \sim \delta b_{\rm rms}/\sqrt{3}$. In the Fourier space, the energy is injected in the $2k_0 \le k \le 7 k_0$ shells, where $k_0=2\pi/L_A=1$ and $k=({k_x^2+k_y^2+k_z^2})^{1/2}$, with a flat energy spectrum. Table \ref{table:MHDruns} summarizes the most relevant numerical parameters of MHD simulations.

\begin{table}
\caption{Parameters of the perturbations used in the MHD simulations.}
\label{table:MHDruns}   
\centering            
\begin{tabular}{c c c c c c} 
\hline   
RUN & $\delta b_{\rm rms}/B_0$ & $\delta v_{\rm rms}/v_A$ &  $\Delta t/t_A$ & $t^*/t_A$\\ 
\hline                      
   A & $0.05$ & $0.05$ &  $10^{-3}$ & $40$ \\
   B & $0.11$ & $0.11$ &  $10^{-3}$ & $16$ \\
   C & $0.23$ & $0.23$ & $10^{-3}$ & $10$ \\
   D & $0.5$ & $0.5$ &  $10^{-3}$ & $3.5$ \\
\hline
\end{tabular}
\end{table}

\subsection{Test-particle solver}
The normalized motion equations of $N_p=10^{3}$ test-protons of mass $m_p$ and charge $e$, i.e.
\begin{eqnarray}
&&\frac{d \br_p}{dt} = {\bm v}_p \label{eq:rsc}\\
&&\frac{d {\bm p}_p}{dt} = \alpha \left( {\bm E} + {\bm v}_p \times {\bm B}   \right) \label{eq:usc}
\end{eqnarray}
are here integrated numerically. In Equations (\ref{eq:rsc}--\ref{eq:usc}) ${\bm r}_p$, ${\bm v}_p$, and ${\bm p}_p=\gamma_p {\bm v}_p$ are the particle position, velocity, and momentum, while ${\bm E}$ and ${\bm B}$ are the electric and magnetic fields generated through the MHD simulations. 
We assume stationary electromagnetic fields and consider a static snapshot of these fields when the turbulence is most developed ($t^*$ in Table \ref{table:MHDruns}, see below). Periodic boundary conditions are also imposed on particle trajectories along each direction.
Equations (\ref{eq:rsc}-\ref{eq:usc}) are scaled analogously to the MHD simulations. In particular, the Lorentz factor is $\gamma = 1/\sqrt{1 - (\beta_A v_p)^2}$, where $\beta_A = v_A/c\simeq 10^{-4}$. Eqs. (\ref{eq:rsc}--\ref{eq:usc}) are integrated by adopting the relativistic Boris method \citep{ripperda2018comprehensive, dundovic2020novel, pezzi2022relativistic}. The electric and magnetic fields are interpolated at the particle position through a trilinear interpolation method \citep{birdsall2004plasma}. The electric field in Equation (\ref{eq:usc}) is the inductive one derived through Ohm's law: ${\bm E}= -{\bm v}\times{\bm B}$. We neglect the resistive electric field for two reasons. First, hyper-viscous terms in MHD simulations, including hyper-resistivity, are not intended to describe any physical effects but only complement the dealiasing procedure in stabilizing numerical simulations. Second, their characteristic scale is much smaller then the minimum particle gyroradius adopted in test-particle simulations, as discussed in the following. 

The parameter $\alpha = t_A \Omega_{0}$, where $\Omega_0=e B_0/m_p c$ is the proton cyclotron frequency, is related to the extension of the turbulent inertial range, since it can be rewritten as $\alpha=L_A/d_p$, where $d_p$ is the proton skin depth of the background plasma \citep{DmitrukEA04,gonzalez2016compressibility}. At variance with \citet{pezzi2022relativistic}, we here consider nonrelativistic particles, with initial speed $\sim v_A$ in a $\beta\sim 1$ plasma. In this case, $1/\alpha$ roughly corresponds to the initial particle gyroradius in normalized units. To require that the particle gyroradius is larger than the gridsize ---thus avoiding spurious numerical effects--- we are forced to artificially reduce $\alpha$. Considering the above normalizing parameters, $\alpha\simeq 50$, i.e. $r_{g} \simeq 0.02 L_A \simeq 2 \Delta x$ and speed $v_p \simeq v_A$. We anticipate that the resonant wavenumber associated with the particle gyroradius $kL_A=L_A/r_g=50$ is close to the end of the inertial-like range of turbulence and larger than either dealiasing scale and dissipative scales at which the resistive electric field is expected to become important. To double-check, we verified that our results in terms of acceleration and transport do not change by including the resistive electric field. Note that the value of the particle gyroradius is limited by setting it to be larger than the grid size. Other methods --based on the stochastic differential equations \citep{wijsen2022observation}-- allow for considering much smaller Larmor radii with respect to the ones here considered through the parametrization of the transport processes, i.e. by prescribing a-priori the particle diffusion coefficient.

When decreasing $\alpha$ to feasible yet unrealistic values such that $r_{g,0}> \Delta x$, we kept small scales ---the ones related to particle motion--- fixed, and we implicitly decreased $L_A$ and the turbulence correlation length $l_c$. In this perspective, the initial particle energy is about few eV (compatible with the thermal plasma population at temperature $T_0$). 
Moreover, all other parameters are similar to solar-wind observations. Such procedure has been already adopted in the literature (e.g., \citet{gonzalez2016compressibility}). 
An alternative, not implemented here, would be to set $L_A$ to realistic values, for example, such that $l_c\simeq 5 \times 10^6 {\rm km}$ and then increase the initial particle gyroradius to values which can be afforded with simulations, i.e., $\alpha\simeq 50$. In this case, one may reach relativistic energies, thus relaxing the constrain on $\alpha$ since for relativistic particles $r_g=1/\alpha$ is not anymore valid \citep{pezzi2022relativistic}. 
Another alternative would consider relativistic electrons (e.g., \citet{trotta2020fast}). In this case, the $\alpha$ factor should include the mass ratio $m_p/m_e$ since MHD is normalized essentially to protons while test-particles would be electrons. 
This issue is in any case not crucial for our study since particle transport and energization are regulated by the relative values of the particle gyroradius $r_g$, the turbulent correlation length $l_c$, and the flux-rope characteristic scales $L_{\rm FR}$ and $L_\Delta$. 
The variability of these parameters, as well as the different options for setting particle energy and species, will be explored in a future work.

\begin{figure}[!htb]
\centering
\includegraphics[width=\columnwidth]{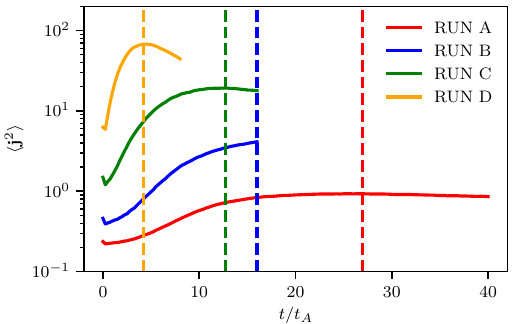}
\caption{Time evolution of $\langle \bj^2\rangle(t)$ for the different runs in Table \ref{table:MHDruns}. Vertical dashed lines indicate the time instant $t^*$ at which, for each run, we perform the subsequent analyses on turbulence and particle transport.}
\label{fig:J2}
\end{figure}

\begin{figure*}[!htb]
 \centering
\begin{minipage}{0.35\textwidth}
\includegraphics[width=\textwidth]{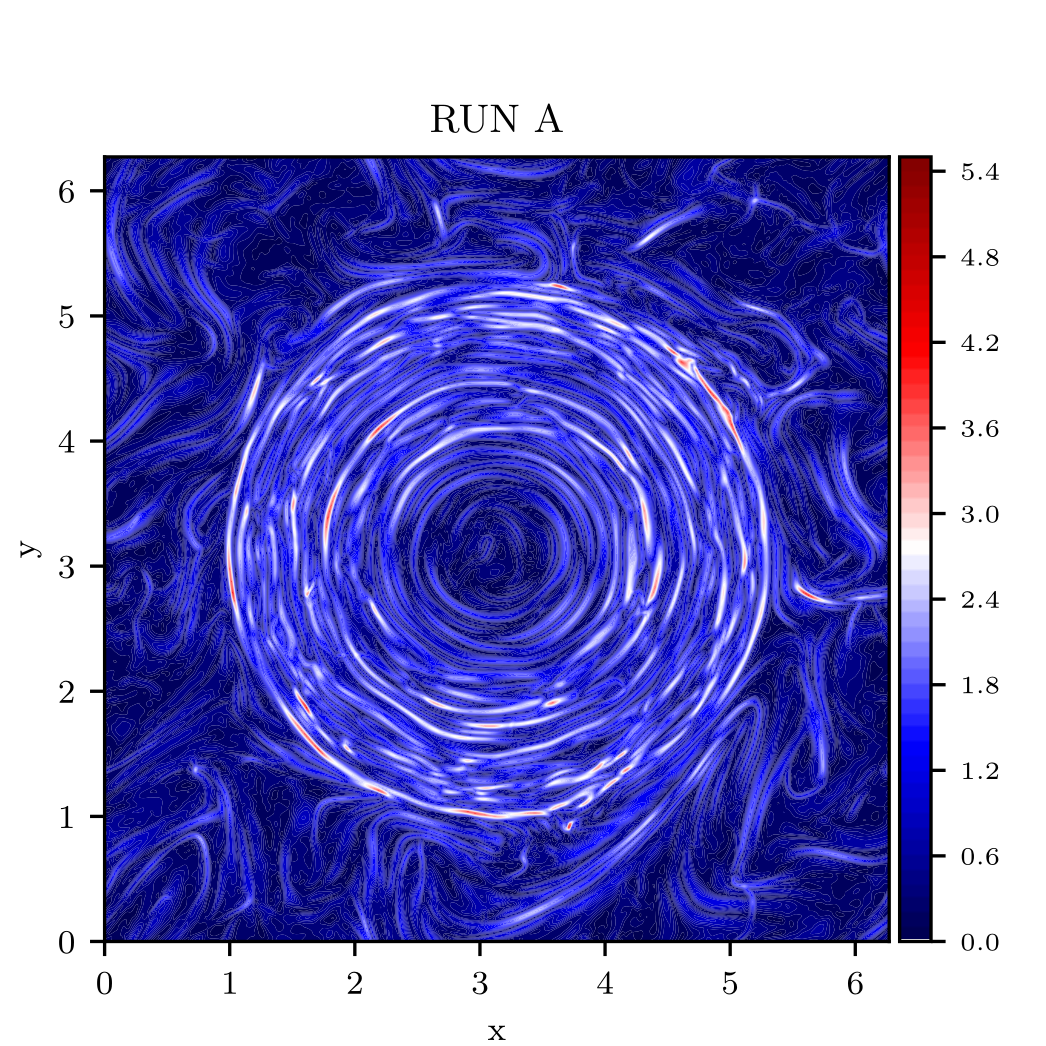}
\end{minipage}
\begin{minipage}{0.35\textwidth}
\includegraphics[width=\textwidth]{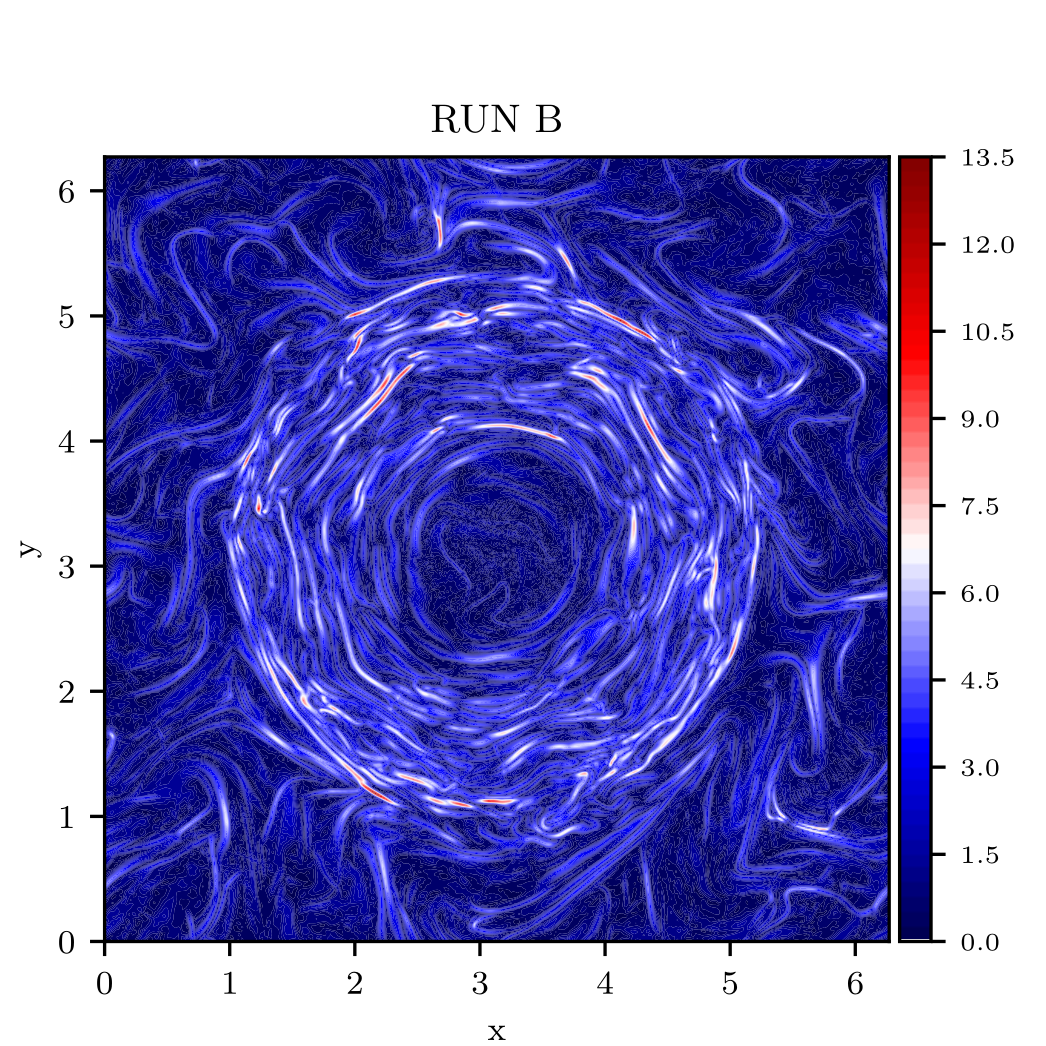}
\end{minipage}
\begin{minipage}{0.35\textwidth}
\includegraphics[width=\textwidth]{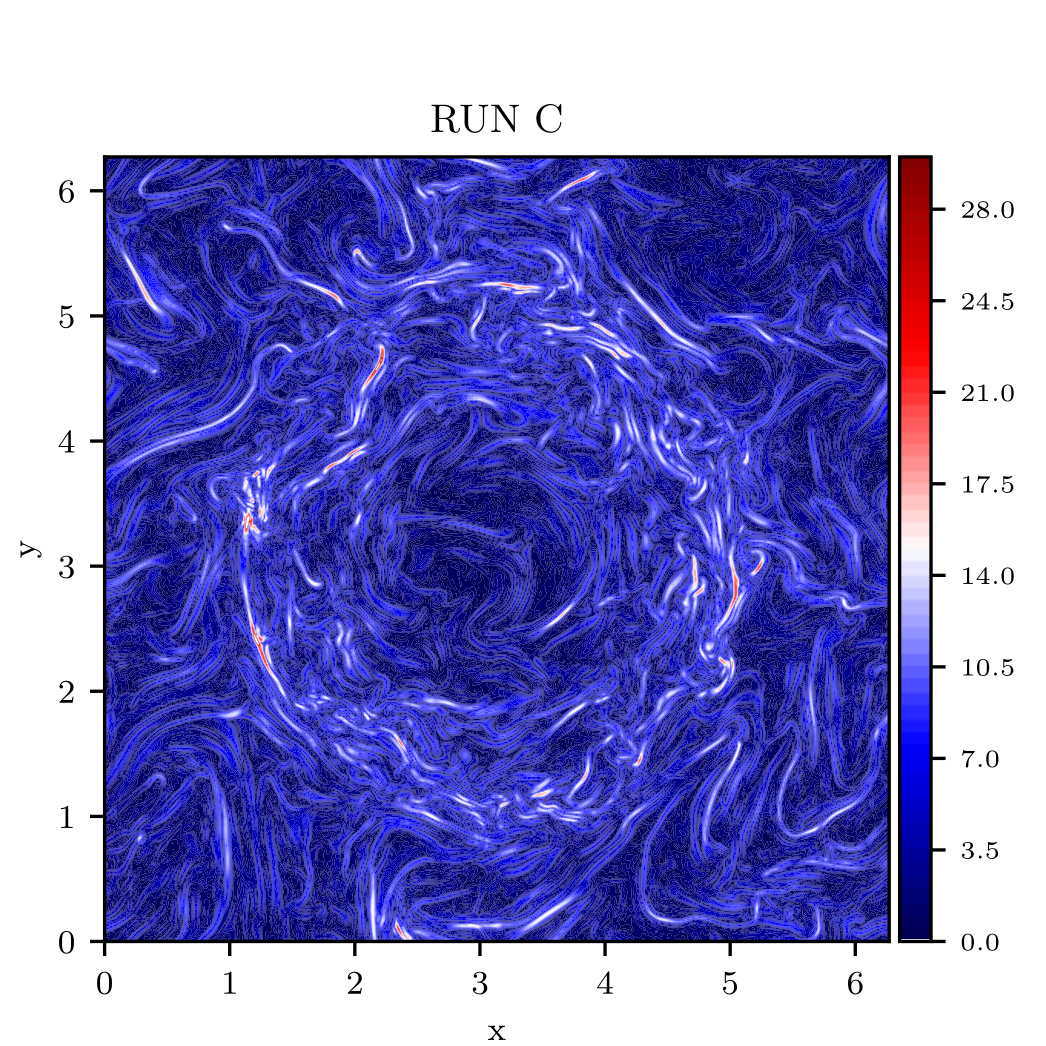}
\end{minipage}
\begin{minipage}{0.35\textwidth}
\includegraphics[width=\textwidth]{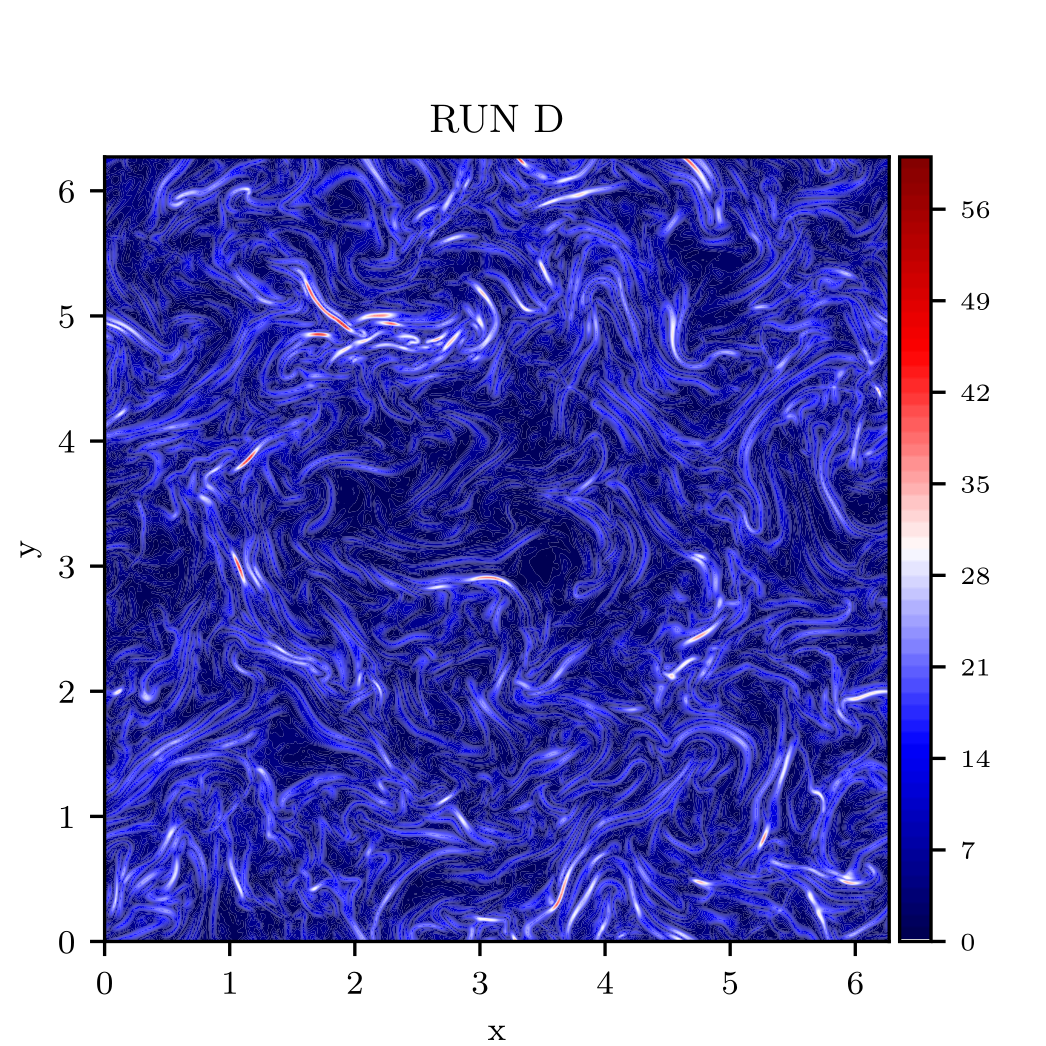}
\end{minipage}
\caption{2D overviews of the current density $|\bj|$ in the plane $(x,y,z=L_z/2)$, perpendicular to the flux rope axis, for the different runs in Table \ref{table:MHDruns} and computed at $t=t^*$. }
\label{fig:jmaps_perp}
\end{figure*}

We performed test-particle simulations by considering different kinds of spatial injections for better understanding the role of the flux rope in particle transport and acceleration. 
In particular, protons are injected: (i) homogeneously throughout the entire 3D computational box; (ii) inside the large-scale flux rope, namely, on the vertical line at $(x,y)=(\pi,\pi)$; (iii) outside the large-scale flux rope, that is on the surface of the open cylinder centered at $(\pi,\pi)$ of radius $r=\pi$.
The first injection is useful to understand global transport properties, while the second and third injections allow us to distinguish regions inside and outside the structure, respectively. 
For all these different cases, initial particle velocities are distributed uniformly on the surface of a sphere of constant energy. 
The numerical time step is always set to $1/20$ of the initial gyroperiod.

\section{Turbulence and intermittency in twisted flux ropes}
\label{sect:turb}

In this section we discuss how turbulence shapes the flux rope and the peculiarities in the energy transfer process directly caused by this large scale structure. 

As a result of the underlying nonlinearities, the energy of perturbations, initially confined at small wave numbers, transfers towards higher wave numbers. A standard proxy of this nonlinear transfer is the time evolution of $\langle {\bm j}^2\rangle(t)$, being $\langle\dots\rangle$ the average over the computational box, which is displayed in Fig. \ref{fig:J2}. In all our numerical experiments, $\langle \bj^2\rangle(t)$ increases to reach a peak, corresponding to the time instant at which turbulent activity is most intense, here denoted with $t^*$. With the exception of Run B where the peak was not fully reached at the end of the run, the peak is then followed by a decreasing phase related to numerical dissipation in these MHD simulations. Hereafter, we focus on the analysis of turbulence at the peak time instant $t^*$.

The distribution of current structures in the plane perpendicular to the flux rope axis, $(x,y,z=L_z/2)$, is depicted in Figure \ref{fig:jmaps_perp} for the different runs in Table \ref{table:MHDruns}. As expected, the different color bar in each panel indicates that stronger initial perturbations are associated with more intense current structures.  

\begin{figure}[!htb]
 \centering
\begin{minipage}{0.35\textwidth}
\includegraphics[width=\textwidth]{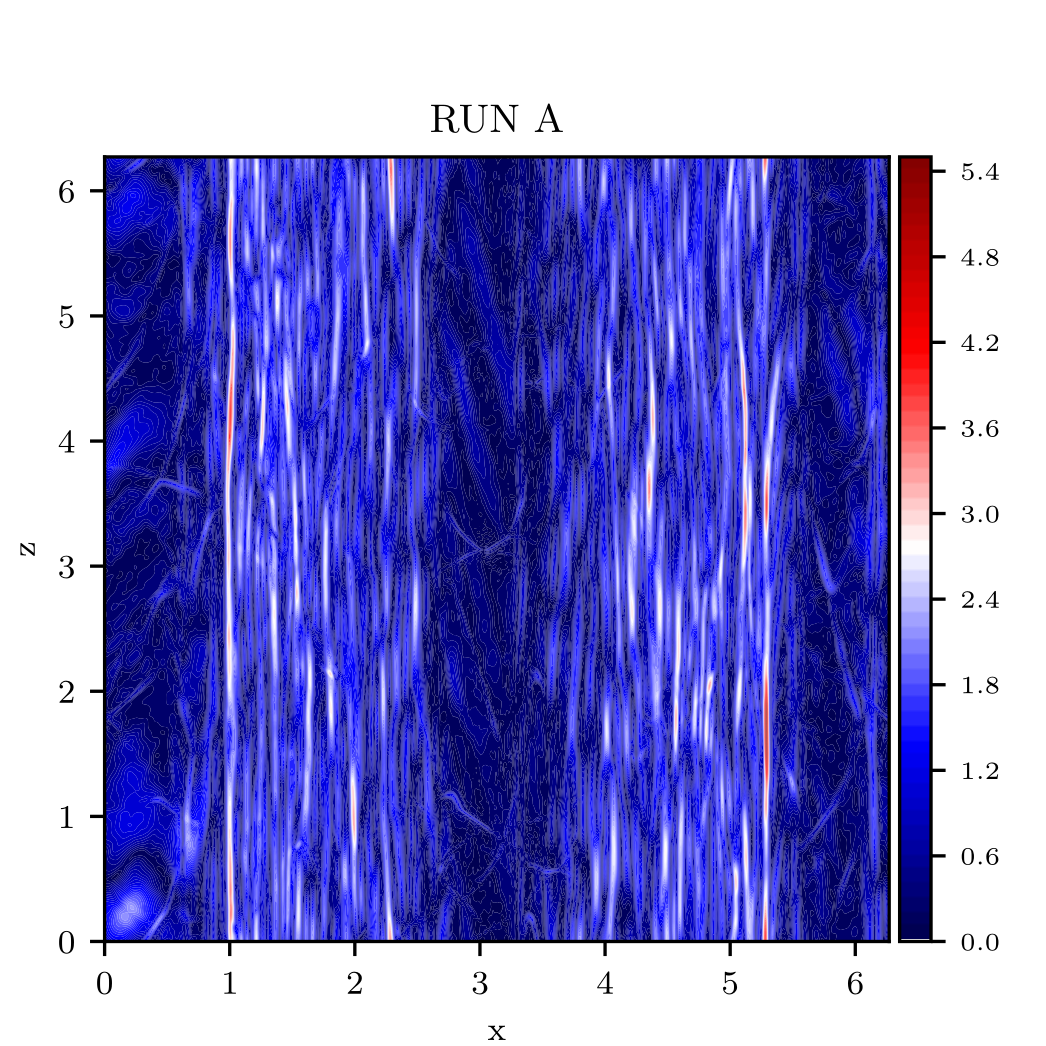}
\end{minipage}
\begin{minipage}{0.35\textwidth}
\includegraphics[width=\textwidth]{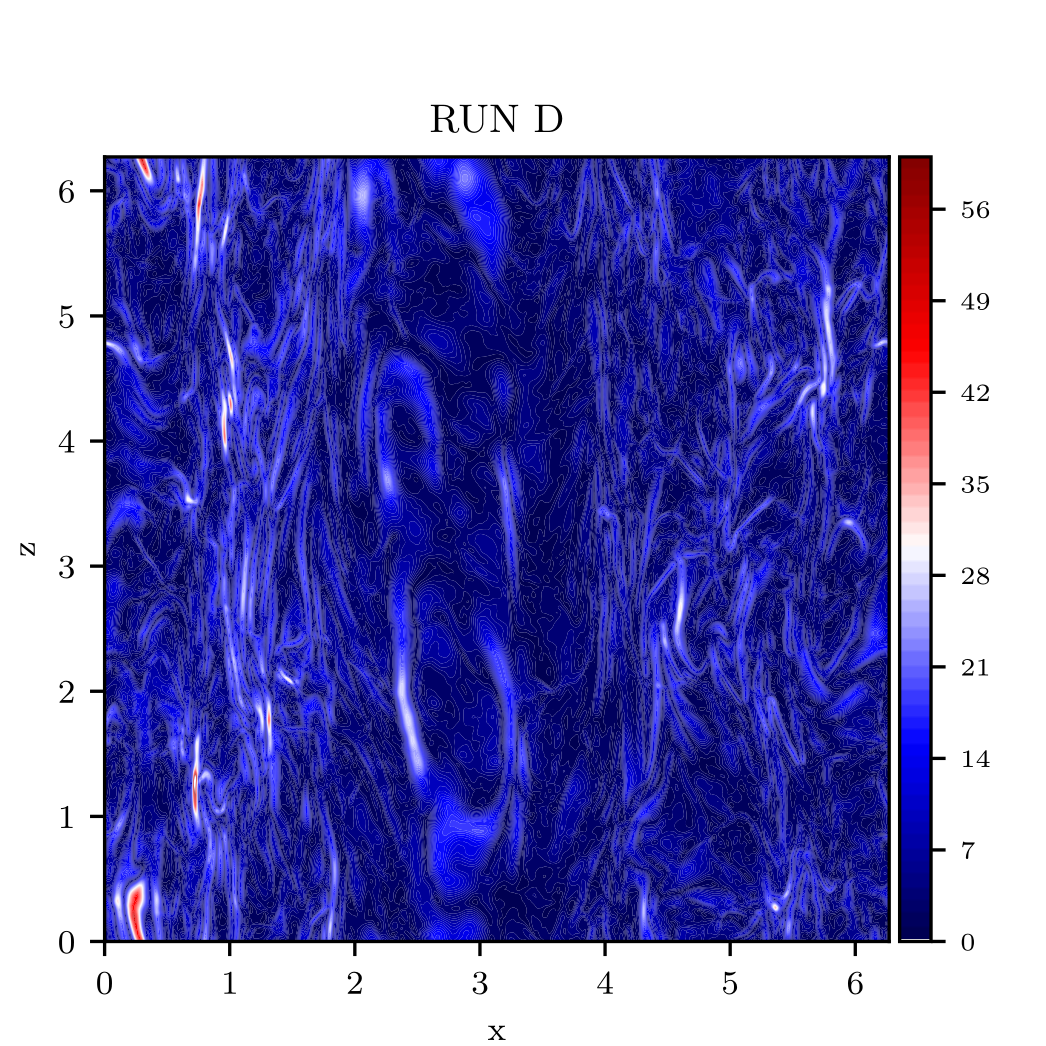}
\end{minipage}
\caption{2D overviews of the current density $|\bj|$ in the plane $(x,y=L_y/2,z)$, parallel to the flux rope axis, for the RUN A (top) and RUN D (bottom) and computed at $t=t^*$. }
\label{fig:jmaps_para}
\end{figure}

For small initial perturbation (top left panel), the most intense current structures are localized along the shear of the flux rope. Erratic current sheets are also observed in the region outside of the flux-rope structure. 
At the shear location, current structures are circular arcs in the plane perpendicular to the flux-rope axis. 
These structures, which may due to phase-mixing along the shear \citep{valentini2017transition, maiorano2020kinetic}, are elongated in the direction of the flux rope axis. 
This behavior, expected from fundamental MHD anisotropy arguments \citep{Shebalin83}, can be appreciated from Fig. \ref{fig:jmaps_para} where the current density maps in the plane parallel to the flux-rope axis, $(x,y=L_y/2,z)$, are displayed for RUN A (top) and RUN D (bottom). 
Multiple anisotropies, induced by both the mean large scale field and the radial magnetic shear, may also arise.

As the level of fluctuations increases, the current structures distribute randomly in the plane perpendicular to the flux rope, resembling the patch of current structures in fully-developed homogeneous turbulence simulations \citep[e.g.,][]{servidio2011magnetic}. 
This feature is confirmed by inspecting the current density maps in the plane parallel to the flux rope axis (Fig. \ref{fig:jmaps_para}, bottom panel). Indeed, in RUN D, current sheets are not localized within the flux-rope boundary and tend to permeate the entire computational plane, diffusing particularly outside the flux rope. 
This suggests that the coupling with the flux-rope shear is less relevant in the presence of strong fluctuations which can self-couple to generate homogeneous-like turbulence. 

It is worth noticing that the core of the flux rope is a region of relatively weak current structures. Such a feature is particularly clear in the case of small initial perturbations. 
This suggests that, far from the shear region, the flux rope inhibits the development of turbulence, and remains --at least in its inner part-- a quasi-equilibrium structure. This behavior can be expected since the flux rope is a MHD equilibrium characterized by a net magnetic helicity, and nonlinear couplings are formally depleted inside it \citep{servidio2008depression}. 

\begin{figure}[!htb]
 \centering
\begin{minipage}{0.35\textwidth}
\includegraphics[width=\textwidth]{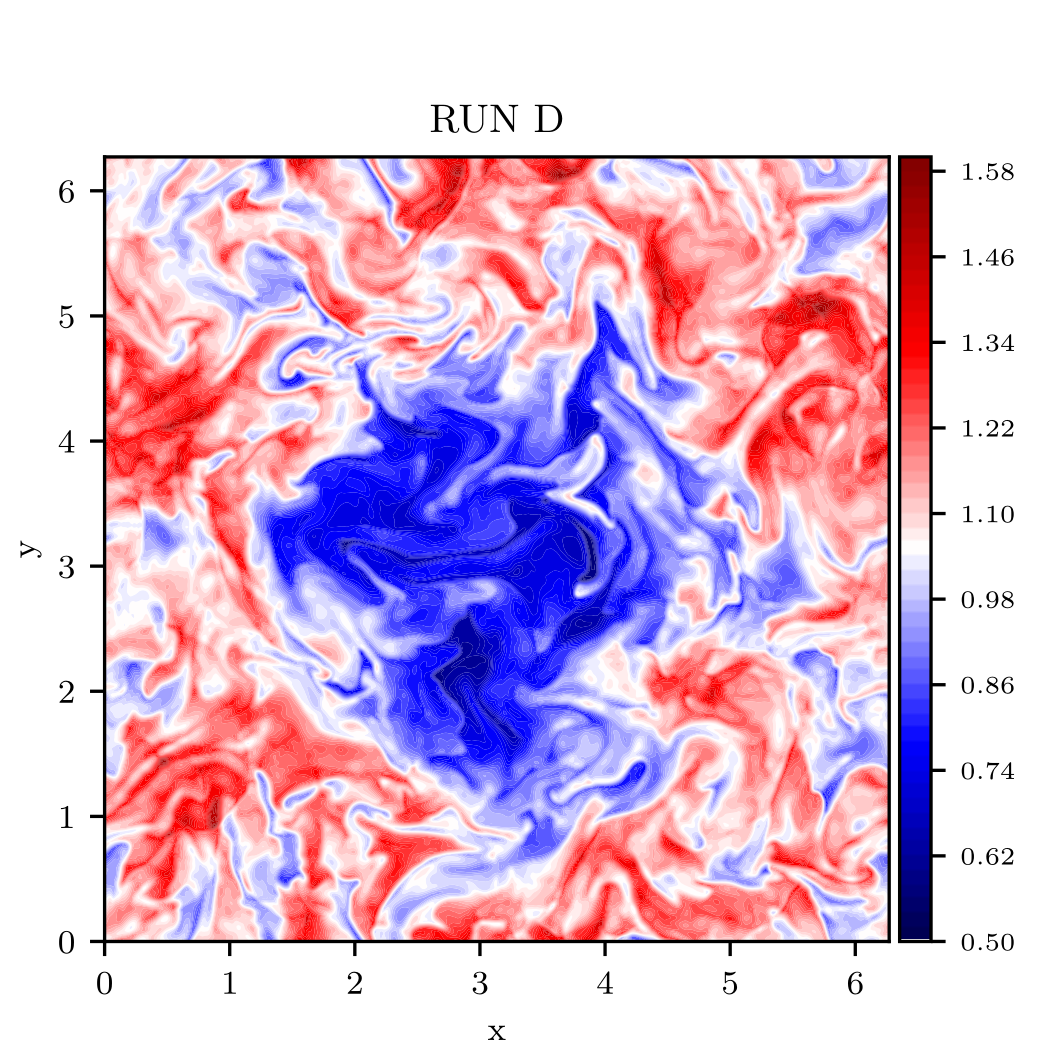}
\end{minipage}
\caption{2D overview of the temperature $T$, at the vertical position $z=L_z/2$, for RUN D and computed at $t=t^*$.}
\label{fig:Tmaps}
\end{figure}

Being a quasi-equilibrium structure, the flux rope is not completely destroyed by fluctuations, even for the strongest perturbation level analyzed here, and despite the presence of small-scale current structures flowing into the entire computational domain. As it can be easily appreciated in Fig. \ref{fig:Tmaps}, which displays, for RUN D, the 2D contour plots of the temperature in the plane perpendicular to the flux-rope axis, the flux rope is still present with its cold core. The temperature pattern is highly-structured outside the flux rope, where warm blobs of plasma on large scales are surrounded by regions with temperature variations at smaller scales, co-located with intense small-scale current sheets (bottom right panel of Fig. \ref{fig:jmaps_perp}). 
These regions also show blobs of cold plasma, possibly transported by turbulent fluctuations from the central part of the flux rope. 
Such transport is suppressed in cases of weak perturbations (not shown here).

\begin{figure}[!ht]
\centering
\includegraphics[width=0.8\columnwidth]{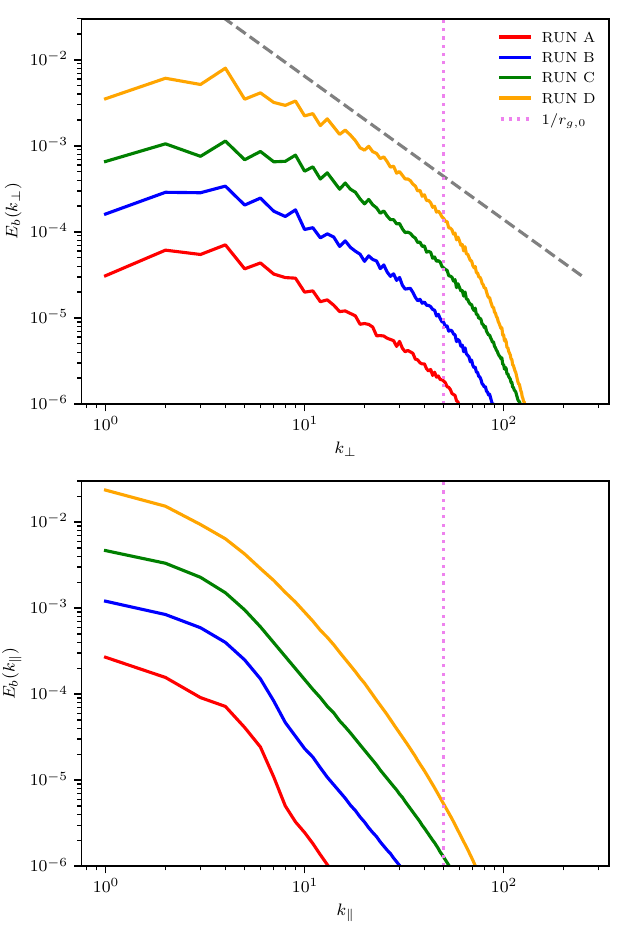}
\caption{Magnetic energy spectra for the different runs in Table \ref{table:MHDruns}, computed at $t=t^*$. Top and bottom plots respectively show the perpendicular and parallel spectra, i.e., $E_b(k_\perp)$ and $E_b(k_\parallel)$. The gray dashed line shows the Kolmogorov prediction $k_\perp^{-5/3}$ \citep{Kolmogorov41a}, as a reference. The violet vertical line indicates the resonant wave number associated with the initial gyroradius $r_{g,0}$ of test particles.}
\label{fig:Spectra_B}
\end{figure}

In order to get further insights about the nonlinear couplings occurring in the system, Figure \ref{fig:Spectra_B} shows the magnetic energy spectra as a function of the wavenumbers perpendicular (top) and parallel (bottom) to the flux rope axis. 
Perpendicular spectra, $E_b(k_\perp)$, are computed by averaging along the parallel direction and assuming isotropy in the perpendicular plane (i.e., summing the energy over circular shells in the $k_\perp$ plane). Similarly, parallel spectra $E_b(k_\parallel)$ are calculated by averaging along the two perpendicular directions in the spectral space. Before computing the magnetic energy spectra, we removed the initial equilibrium magnetic field, whose gradients are nevertheless confined at low wavenumbers.

As a result of nonlinear couplings, the energy of fluctuations, initially confined at large scales, spreads towards higher wave numbers. 
The energy-containing scale of fluctuations, visually estimated as the location of the spectral peak in Fig. \ref{fig:Spectra_B}, is roughly $l_c\sim 1-1.5 L_A$. 
Velocity spectra are similar to magnetic ones, while compressive effects remain, on average, weak. 
The transfer of turbulent fluctuations is anisotropic, and parallel spectra are generally steeper than perpendicular ones. 
In the perpendicular directions, a power-law spectrum, with a slope compatible with the Kolmogorov prediction, is generated and extends for about a decade ($6 \lesssim k_\perp \lesssim 50$). 
At higher wavenumbers, numerical dissipative effects steepen the spectrum. 
Interestingly, such a Kolmogorov-like spectrum forms also when initial fluctuations are weak and the nonlinear coupling is also affected by the coupling of the perturbations and the flux-rope shears via, e.g., phase-mixing.

\begin{figure*}[!t]
\centering
\includegraphics[width=0.7\textwidth]{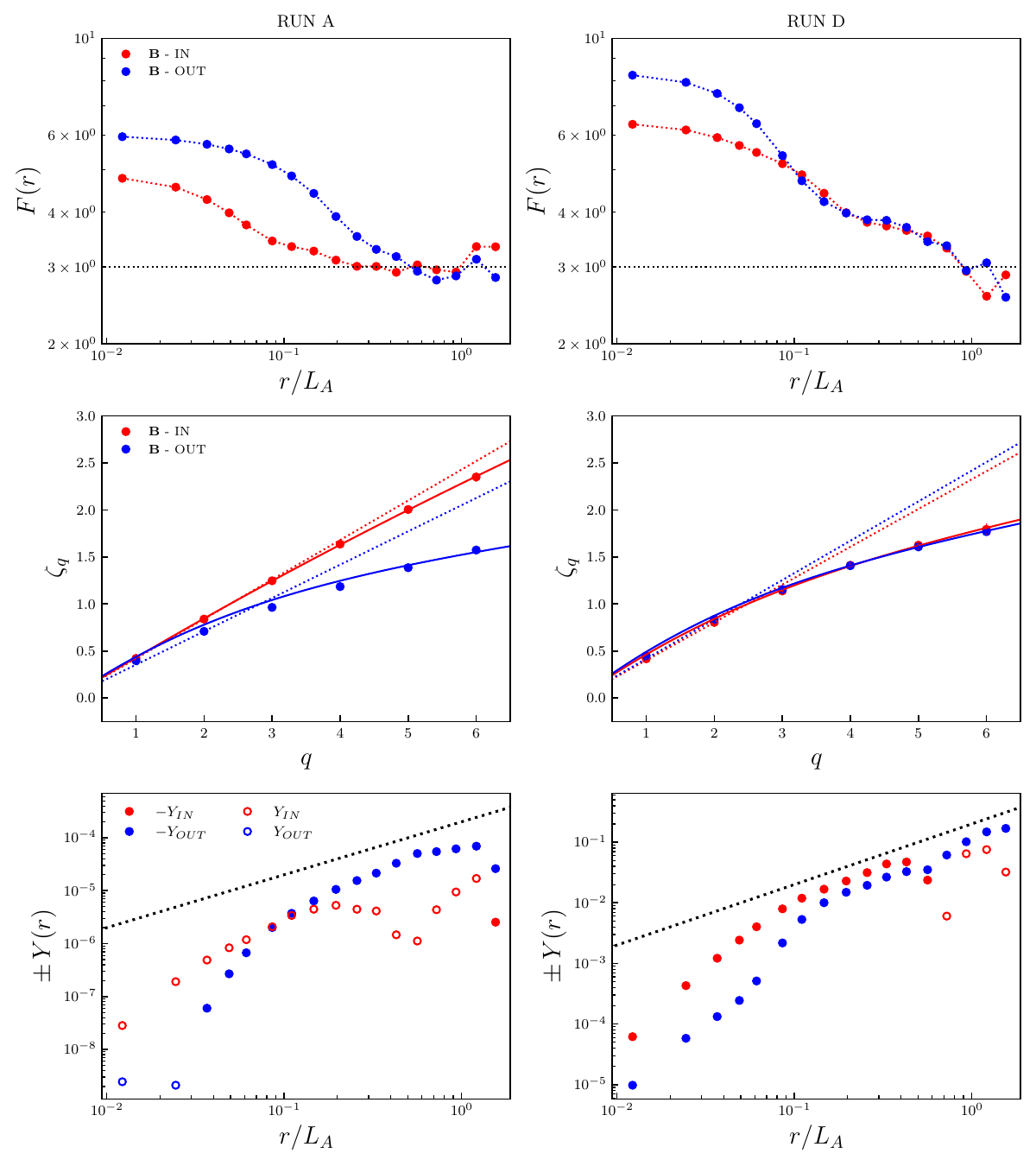}
\caption{Top: Flatness $F(r)$ of the magnetic field increments as a function of the separation scale $r$ for RUN A (left) and RUN D (right) and for data sampled inside (red) and outside (blue) the flux rope. Dotted line is the reference Gaussian flatness value. Middle: Scaling exponents $\zeta_q$ as a function of the order $q$ for RUN A (left) and RUN D (right) evaluated inside (red) and outside (blue) the flux rope for the magnetic field. Dotted lines represent the linear scaling, while solid lines indicates the $p$-model best fit for each data sample. The Hurst exponents $h=\zeta_2/2$ are 0.42 (inside) and 0.35 (outside) for RUN A, and 0.40 (inside) and 0.42 (outside) for RUN D. Bottom: Politano-Pouquet's term $Y(r)$ as a function of the separation scale $r$, computed inside (red) and outside (blue) the flux rope, for RUN A (left) and RUN D (right). The dotted lines indicate the linear scaling. Filled circles correspond to $-Y(r)$ and empty circles to $Y(r)$.}
\label{fig:zetapscalings}
\end{figure*}

We complement the above-discussed global spectral analysis with a local analysis, based on the increments of the magnetic field $\bb$ and the flow speed $\bv$ and aimed at highlighting differences occurring in turbulent transfer inside and outside the flux rope. 
Given the 2.5D configuration of the equilibrium flux rope, and considered that the most intense nonlinear couplings occur in the directions transverse to the flux rope axis, the following analysis has been carried out in the $(x,y)$ plane. 
In order to improve the statistics, we performed an ensemble average over all the $N_z=512$ planes perpendicular to the flux-rope axis. For consistency with the magnetic energy spectra, we removed the initial equilibrium structure. 
Starting from the entire plane $(x,y)$, we extracted four straight cuts of length $L/2=\pi L_A$ along the $x$ and $y$ directions. Two of them are centered in the equilibrium structure and are representative of the inner part of the flux rope. Another two are instead placed at the boundaries of the simulation domain and probe the fields outside the flux rope. On each of these cuts, we separately computed the longitudinal increments of the magnetic field, flow speed, and Elsasser variables. 
Finally, the two set of increments have been combined in the analysis for improving the statistics.

We first calculated the structure functions of order $q$ as $S^{(q)}_r=\langle \Delta B_r^q\rangle$, where $\Delta B$ indicates the generic longitudinal increment of the magnetic field at the spatial scale $r$. 
The structure functions are expected to grow as power laws of the increments' scale separation \citep{Frisch95}. 
Due to the intermittent nature of turbulence, which is characterized by intense inhomogeneities in both space and time at different scales, the scaling exponents $\zeta_q$ of the structure functions, $S^{(q)}_r\sim r^\zeta_q$, exhibit an anomalous scaling as a function of the order $q$, which deviates from the linear trend predicted by globally self-similar theories of turbulence \citep{Kolmogorov41a,Kraichnan65}. Consequently, the flatness of magnetic field increments, defined as $F(r)\equiv S_r^{(4)}/[S_r^{(2)}]^2$, exhibits an increasing power-law trend from large to small scales \citep{Frisch95,Carbone2014}. 

The top panels of Figure \ref{fig:zetapscalings} display the flatness $F(r)$ of the magnetic field increments as a function of the scale separation $r$, obtained inside (red) and outside (blue) the flux rope for RUN A (left) and RUN D (right). Outside the flux rope, $F(r)$ exhibits a similar power-law increase in the inertial range ($r\gtrsim 0.2-0.3 L_A$) in both RUN A and RUN D. Inside the flux rope, the profile of $F(r)$ is similar to what found outside the flux rope for RUN D. In contrast, the flatness $F(r)$ does not exhibit the typical power-law range in the inertial scales for RUN A. 
These features suggest that the fluctuations are intermittent outside the flux rope, with properties qualitative independent of the initial perturbations amplitude. On the other hand, similar intermittency is observed inside the flux rope only for large amplitude perturbations.

To further investigate this aspect, we show in the middle panels of Figure \ref{fig:zetapscalings} the scaling exponents $\zeta_q$ as a function of the order $q$, obtained inside (red) and outside (blue) the flux rope for RUN A (left) and RUN D (right). Given the limited range of scales of the simulation, we adopted the extended self-similarity method \citep{benzi1993extended} for estimating scaling exponents more reliably. 
In both RUN A and RUN D, the magnetic field scaling exponents computed outside the flux rope (circles) deviate from the linear scaling $\zeta_q\sim hq$, where $h=\zeta_2/2$ indicates the Hurst exponent, predicted by assuming the global self-similarity (dotted lines in Fig. \ref{fig:zetapscalings}). 
For reproducing the observed anomalous scaling trend, we performed the best fit of $\zeta_q (q)$ adopting the following expression obtained from the $p$-model \citep{Meneveau87} (solid lines in bottom panels of Fig. \ref{fig:zetapscalings}):
\begin{equation}
    \zeta_q=1-\log_2[p^{hq}+(1-p)^{hq}].
\end{equation}
The parameter $p$ of the model, which represents the weight of the energy repartition in the multiplicative process used to mimic the turbulent energy cascade, is $p\simeq 0.82$ in RUN A outside the flux rope, and $p\simeq 0.78$ in RUN D both outside and inside. 
Such values indicate strong intermittency, larger than typically observed in neutral flows \citep{Meneveau87}, and roughly compatible with estimates in the solar wind \citep{carbone1993, sorriso2018local} and in the terrestrial magnetosheath \citep{Yordanova08,Quijia2021}. Conversely, for small initial perturbations (RUN A) the magnetic field's scaling exponents exhibit a quasi-linear scaling inside the flux rope. 
A quasi-linear trend of this kind can be associated with the interaction between turbulence and large-scale gradients of the flux rope, which in turn tend to inhibit the complete development of turbulence for low amplitudes of perturbations. 
As the level of initial perturbations increases (RUN D), indeed, the values obtained for $\zeta_q$ inside the flux rope reconciles with the one observed outside in the case of strong initial perturbations. 

Finally, we complemented the above framework focusing on the properties of intermittency by assessing how the nonlinear energy transfer typical of the inertial range of scales is influenced by the presence of the large-scale flux rope. 
To this purpose, we use the ensemble of cuts defined above to estimate the Politano-Pouquet law for homogeneous, locally isotropic, incompressible MHD turbulence \citep{politano1998vonkarman, carbone2009on}. 
Defining $Y^\pm=\langle|\Delta\bm{z}^\pm|^2 \Delta z^\mp_\parallel\rangle$ as the mixed third-order structure functions of the Elsasser variables $\bm{z}^\pm=\bm{v}\pm (4\pi\rho)^{-1/2} \bm{b}$, the Politano-Pouquet law predicts 
\begin{equation}
    Y(r) = Y^+(r)+Y^-(r) = -\frac{4}{3}\epsilon r,
\end{equation}
where $\epsilon$ indicates the total energy transfer rate of the turbulent cascade.

The bottom panels of Figure \ref{fig:zetapscalings} show the terms $-Y(r)$ (filled circles), obtained for RUN A (left) and RUN D (right) and computed inside (red) and outside (blue) the flux rope. When $-Y(r)$ exhibits a linear scaling, an energy transfer towards smaller scales occurs. Figure \ref{fig:zetapscalings} also presents the values obtained for $Y(r)$ (empty circles), which, conversely, can be associated with an inverse energy cascade \citep{sorriso2007observation,marino2023scaling}. As illustrated in the bottom panels of Figure \ref{fig:zetapscalings}, outside the flux rope for RUN A and both outside and inside the flux rope for RUN D, a rather pronounced linear regime tends to emerge in the inertial range, indicative of the occurrence of a direct turbulent energy transfer. 
In contrast, inside the flux rope for RUN A, $Y(r)$ is always positive, thus suggesting the absence of energy transfer towards smaller scales. 
Moreover, no linear trends are observed in the range of scales corresponding to the inertial range outside the flux rope. Interestingly, when computing the Politano-Pouquet law inside the flux rope, a characteristic scale emerge, $r^*\sim L_A$, at which the scaling law breaks down, for both RUN A and RUN D. 
This could highlight both the energy-containing scale of turbulence $l_c\simeq 1-1.5L_A$ or the coupling of the fluctuations with the gradients of the background equilibrium, whose typical size is again $L_\Delta\simeq L_A$. Indeed the inhomogeneous equilibrium flux rope may dynamically influence the properties of the nonlinear energy transfer \citep{wan2009thirdorder}, and the simple removal of the equilibrium structures before the statistical analysis does not exclude its possible dynamical effects on the cascade.
By extensively testing different regions outside the flux rope, we finally observed that subtle variations in the emergence of a stable linear scaling in the Politano-Pouquet law may occur (not shown). These issues may be related to both the limited scale separation between the large-scale equilibrium flux rope and the turbulent fluctuations, and the limited box size which influences the length of the cut (here of the order of some correlation lengths).

To summarize this Section, the analysis of intermittency and energy transfer rate confirms that the flux rope generally inhibits the nonlinear transfer of turbulent energy across scales when turbulence is relatively weak. In this case, other processes, including linear phase-mixing, are responsible for transferring energy towards smaller scales in the shear region of the flux rope. Otherwise, in regions outside the flux rope, as well as inside of it for large amplitude perturbations, the turbulence exhibits consistent statistical features in terms of scaling laws and energy transfer rate. 

\section{Particle transport and energization in twisted flux ropes}
\label{sect:acc}

\begin{figure}[!b]
 \centering
\begin{minipage}{0.8\columnwidth}
\includegraphics[width=\textwidth]{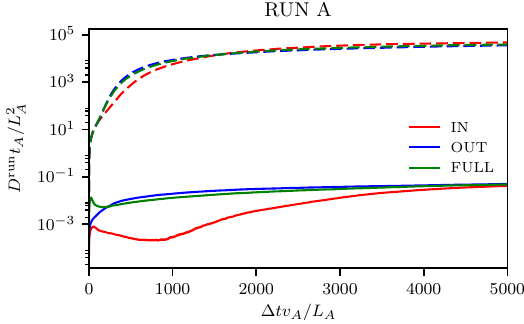}
\end{minipage}
\begin{minipage}{0.8\columnwidth}
\includegraphics[width=\textwidth]{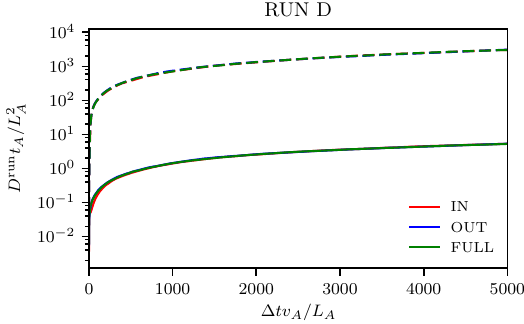}
\end{minipage}
\caption{Time evolution of the parallel $D^{\rm run}_\parallel(\Delta t)$ (dashed lines) and perpendicular $D^{\rm run}_\perp (\Delta t)$ (solid lines) running diffusion coefficients for RUN A (top) and RUN D (bottom).}
\label{fig:Drun}
\end{figure}

In this Section, we explore how the presence of the turbulent flux rope influences particle transport and acceleration. 
We focus on test-particle simulations where the static electromagnetic fields are selected from the MHD simulations at $t=t^*$. 
Assuming static fields implies that the characteristic times of particle transport and acceleration are smaller compared to the times responsible for the formation and possible dissipation of the underlying flux rope. However, as discussed in the following, our results indicate that diffusive and the slowest acceleration times are actually about $\sim 10^3 t_A$, while the perturbed flux rope has been generated within a smaller time, i.e. $t^*\sim 50-100 t_A$. 
Such a finding would suggest to perform test-particle simulations into non-static fields \citep{Gonzalez2017, Trotta2019}. 
We leave this aspect for future works and justify the static assumption as follows. 
First of all, being quasi-equilibrium structures, it is not unreasonable to assume that flux ropes are long-lived structures, capable of traveling in the solar wind over times much longer than the standard energy transfer time of fully-developed turbulence, the latter being a rough estimate of the dynamical time associated with the turbulent flux rope. 
Moreover, our simulations are in decay, while in the solar wind one could imagine a continuous injection of energy at large scales responsible for generating perturbed flux ropes statistically similar to the one adopted in our test-particle simulations. 
In such a case, even if dissipated, these perturbed flux ropes may be quickly generated again by plasma turbulence on short time scales, thus being available to efficiently contribute to particle transport and acceleration. 
Finally, acceleration processes occur with different characteristic timescales and the fast energization process, relevant for efficiently energizing trapped particles, has a characteristic time smaller than or comparable with the dynamical time associated with the turbulent flux rope.

Our analysis begins by exploring the properties of particle transport in physical space. 
For each type of particle injection as described in Sec. \ref{sect:nummod}, we calculated the running time diffusion coefficients along each direction as $D^{\rm run}_{\rm ii}(\Delta t) = \langle (\Delta r_{p,i}(\Delta t))^2 \rangle/2 \Delta t$, being $\Delta r_{p,i}(\Delta t)$ the particle displacement along the $i$-th direction ($i=x,y,z$) during a time interval $\Delta t=t-t_0$ ($t_0=0$), and $\langle ... \rangle$ the average on the particle ensemble. 
Figure \ref{fig:Drun} shows the time evolution of the running time diffusion coefficients perpendicular and parallel to the background magnetic field (i.e., to the flux-rope axis), respectively $D^{\rm run}_\perp(\Delta t)=(D^{\rm run}_{\rm xx}(\Delta t)+D^{\rm run}_{\rm yy}(\Delta t))/2$ and $D^{\rm run}_{\parallel}(\Delta t)=D^{\rm run}_{\rm zz}(\Delta t)$. 
For simplicity we focus on the two extreme perturbations levels, RUN A (top) and RUN D (bottom). 

On short time lags $\Delta t$, interesting differences arise in $D^{\rm run}_\perp$ for small amplitudes of the initial perturbations (RUN A). 
When injecting particles inside the flux rope (red lines in Fig. \ref{fig:Drun}), the perpendicular diffusion coefficient is significantly smaller than the corresponding values achieved injecting particles throughout the entire box or outside the flux rope (blue and green lines in Fig. \ref{fig:Drun}). 
We interpret this behavior as a trapping effect of the flux rope, hindering perpendicular diffusion. The particle escape from the flux rope, where perpendicular diffusion is restored, is probably due to the underlying acceleration process occurring within the flux rope, which prevents trapping after reaching sufficiently large particle gyroradii. 
As the time lag $\Delta t$ increases, these differences disappear and running diffusion coefficients, which do not depend on the kind of particle injection, tend to reach a diffusive plateau in all cases. 
Since particles continue to be stochastically accelerated also on long times, the running diffusion coefficients do not formally achieve the diffusive plateau.

\begin{figure}[!htb]
 \centering
\begin{minipage}{0.35\textwidth}
\includegraphics[width=\textwidth]{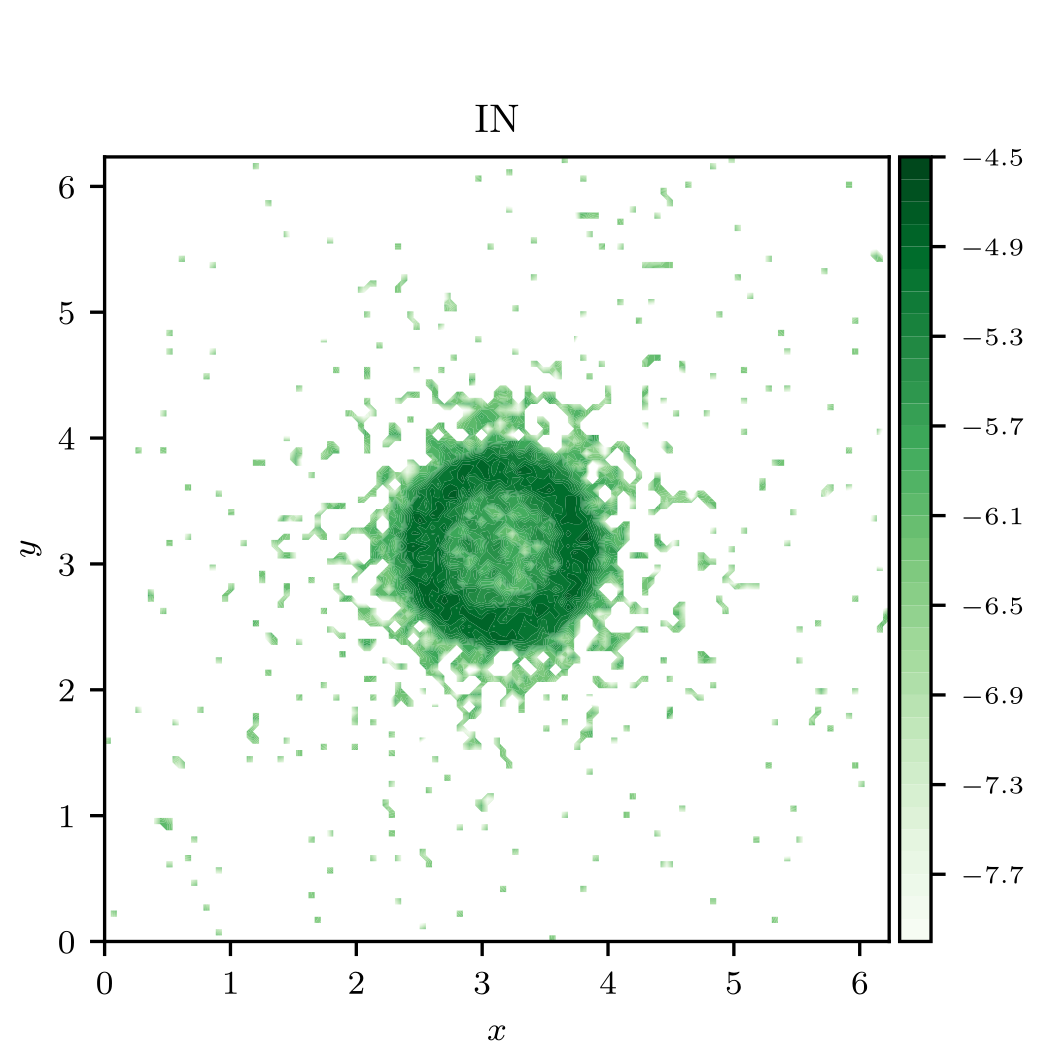}
\end{minipage}
\begin{minipage}{0.35\textwidth}
\includegraphics[width=\textwidth]{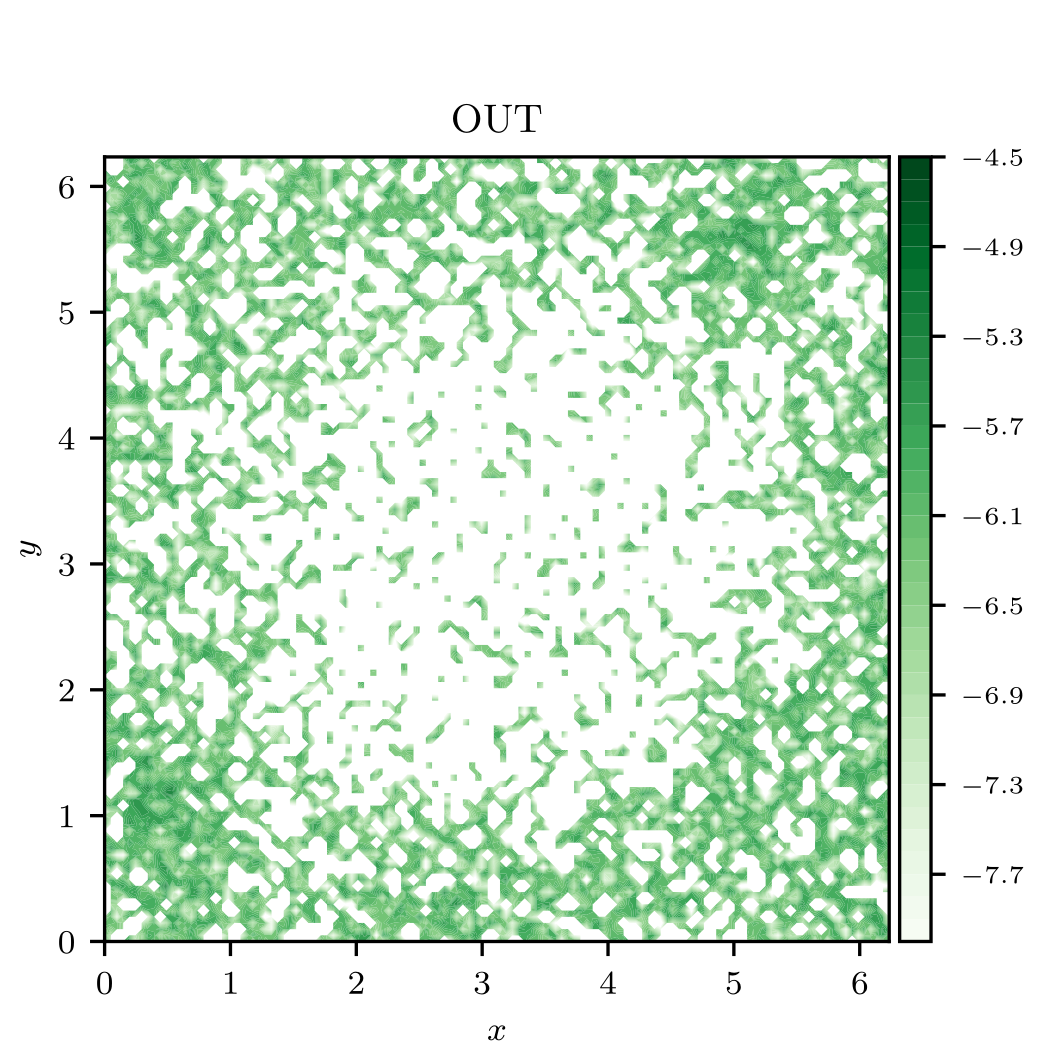}
\end{minipage}
\begin{minipage}{0.35\textwidth}
\includegraphics[width=\textwidth]{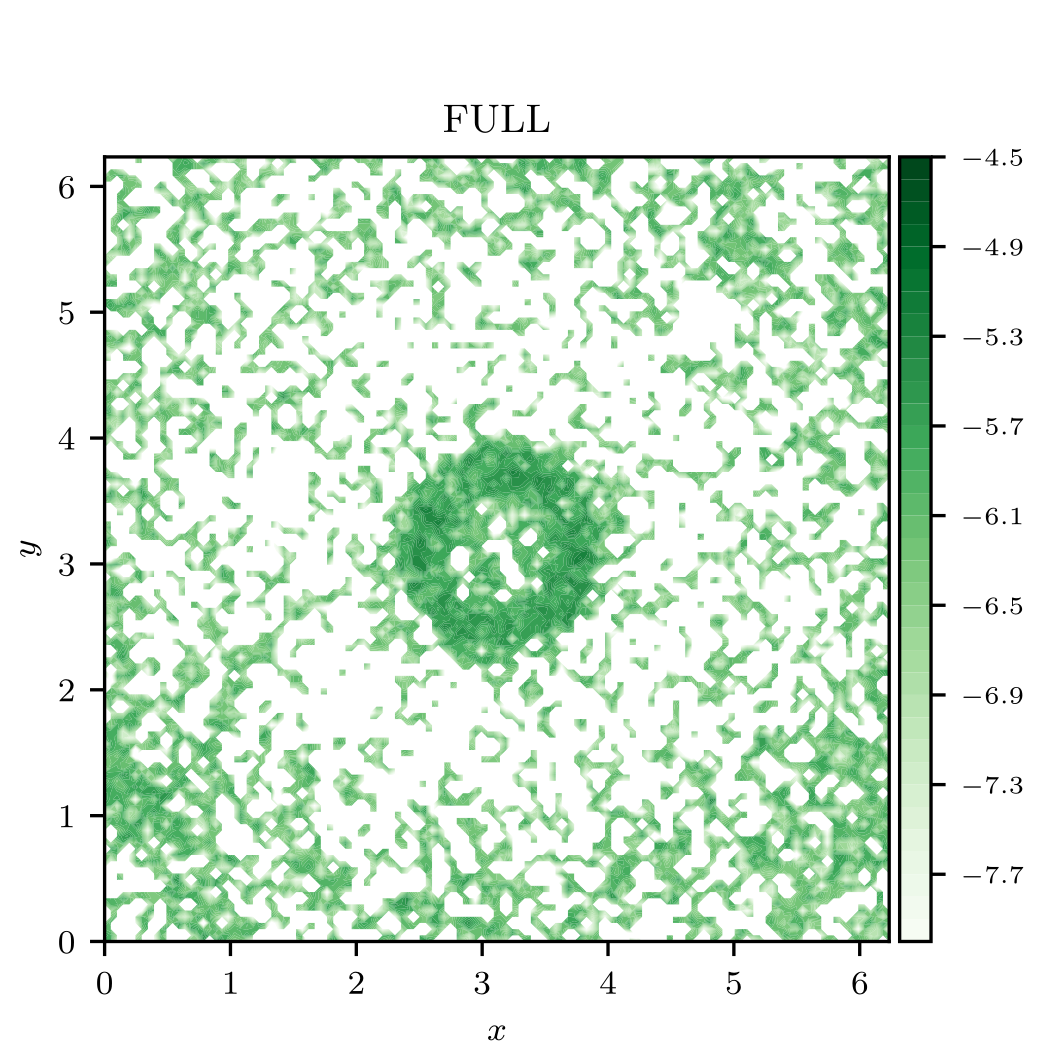}
\end{minipage}
\caption{Particle density in the perpendicular plane $(x,y)$, averaged along the $z$ direction, for the RUN A and at the time $t\simeq1200 t_A$. Top, center, and bottom plots refer to injecting particles inside, outside the flux rope, and randomly in the entire box, respectively.} 
\label{fig:Density}
\end{figure}

To highlight that particles are actually trapped within the large-scale flux rope in the case of weak initial perturbations (RUN A), Figure \ref{fig:Density} shows the particle density in the plane perpendicular to the flux rope axis for the injection inside (top), outside (center) the flux rope and throughout the entire box (bottom). We select the time instant $t\simeq 1200 t_A$, corresponding to the phase in which the perpendicular running diffusion coefficient is weaker when injecting particles inside the flux rope. 
Figure \ref{fig:Density} shows that, in the case of injection inside the structure, particles condensate inside the flux rope, thus limiting the perpendicular transport. Conversely, when injecting particles outside the flux rope, particles preferentially stay outside the structure, though some particles enter it. 
Finally, when injected throughout the entire box, a combination of the two effect is observed with a roughly ergodic particle arrangement in space, with a region of higher density in the flux rope.

\begin{figure}[!htb]
 \centering
\includegraphics[width=\columnwidth]{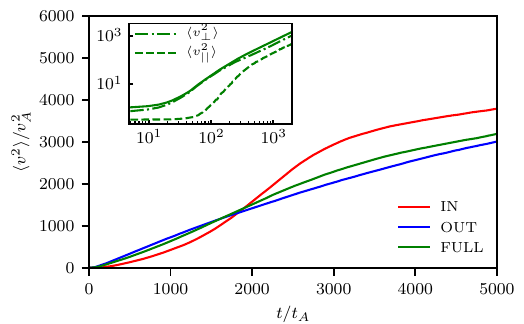}
\caption{Average kinetic energy as a function of time for the RUN A. 
The inset shows, for the case of uniform injection, the time evolution of the average perpendicular (dot-dashed) and parallel (dashed) energies. For reference, we also report the average kinetic energy in solid line. 
}
\label{fig:Enav}
\end{figure}

We then explore the properties of particle acceleration in the perturbed flux rope. 
In particular, Figure \ref{fig:Enav} displays the time evolution of the average kinetic energy for the different injections considered in this work for RUN A. 
We focus here on this low-amplitude run, where the influence of the flux rope on particle dynamics is the most significant. 

When injecting particles outside the flux rope (blue line) or in the entire computational domain (green line), the average particle energization displays a linear growth, possibly characterized by two distinct regimes --the first one, which occurs for $t\lesssim 3000 t_A$, being faster than the second. Conversely, when particles are released inside the flux rope, the average energization is exponential-like in the first time window ($t\lesssim 3000 t_A$), while later it shows a linear trend, whose slope is similar to the one achieved with the other kinds of particle injections in this late time range. 

The energy corresponding to the change from the (fast) exponential growth to the linear one in the case of particles injected in the flux rope roughly corresponds to a particle gyroradius $r_g \sim L_{\rm \Delta}$. Indeed, since $r_{g,0}=0.02 L_A$ and $v_{p,0}\simeq v_A$, the condition $r_g\simeq L_{\rm \Delta} \simeq L_A$ yields $v \simeq 50 v_A$, i.e. $\langle v^2 \rangle \simeq 2.5 \times 10^3 v_A^2$. 
This suggests that the mechanisms responsible for the fast acceleration preferentially observed when injecting particles inside the flux rope are, on average, no longer efficient when $r_g > L_{\rm \Delta}$. 
These mechanisms could be presumably active at the flux-rope boundary, thus being maximized when injecting particles in the flux rope, and be associated with particles trapped in the magnetic structures and experiencing an intense electric field therein, essentially due to gradients in the plasma bulk speed \citep{pezzi2022relativistic}. 
To support this interpretation, we verified that RUN B is still characterized by a double regime of acceleration. Similarly to RUN A, the first phase of energization is exponential-like and breaks at a value of the average particle energy compatible with $r_g \simeq L_{\rm \Delta}$. However, the acceleration is faster in RUN B compared to RUN A. Hence, the characteristic acceleration timescale decreases as the intensity of turbulence increases. Given the presence of the equilibrium magnetic structure with no bulk speed counterparts, the first-order electric field is $E \propto \delta v B_0$, where $\delta v$ is the bulk speed perturbation and $B_0$ is the equilibrium, inhomogeneous magnetic field. 
This rough estimate of the inductive electric field indicates that stronger turbulence is associated with more intense electric fields that accelerate particles, thus implying a faster (on average) process of energization. 

\begin{figure}[!htb]
 \centering
\includegraphics[width=0.8\columnwidth]{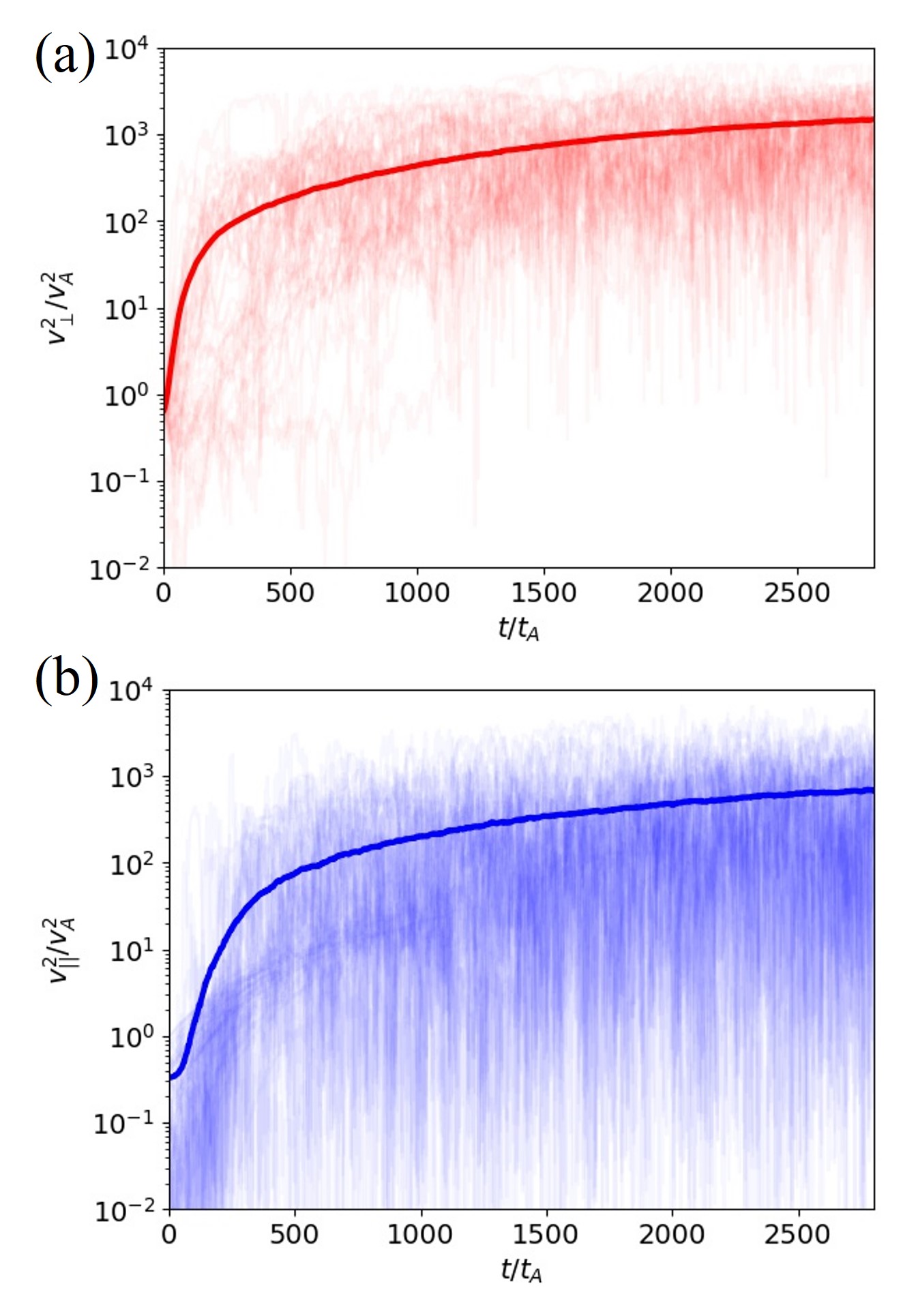}
\caption{Top: Perpendicular kinetic energy as a function of time for averaged over the whole particle ensemble (solid line) and for individual particles (shaded lines) for the case of uniform injection. Bottom: Parallel energy, presented in the same fashion as above.}
\label{fig:par_perp}
\end{figure}

When injecting particles outside the flux rope or throughout the entire box, the average particle acceleration still exhibits a double phase, characterized by a fast, yet linear, initial acceleration. This indicates that less particles compared to the case of injection inside the flux rope still experience the electric field responsible for fast acceleration, thus implying the steepening of the average energization process. In other words, a particle injected outside the flux rope has a finite yet small probability to enter the flux rope. Note that, for stronger initial perturbations (RUN C and RUN D, not shown), the energization is similar for all the different injections. This confirms that the intense turbulent fluctuations mask the effect of the large-scale flux rope in accelerating particles. 

Therefore, turbulent fluctuations have a threefold impact on particle energization. They in general allow for particle acceleration since the background flux rope is purely magnetic (no electric field). Below some threshold, turbulence tends to decrease the characteristic times associated with fast exponential acceleration of trapped particles. Above the threshold, turbulent fluctuations mask the effect of the underlying flux rope, thus producing a detrapping effect on particles possibly confined in the flux rope.

\begin{figure*}[!htb]
 \centering
\includegraphics[width=0.7\textwidth]{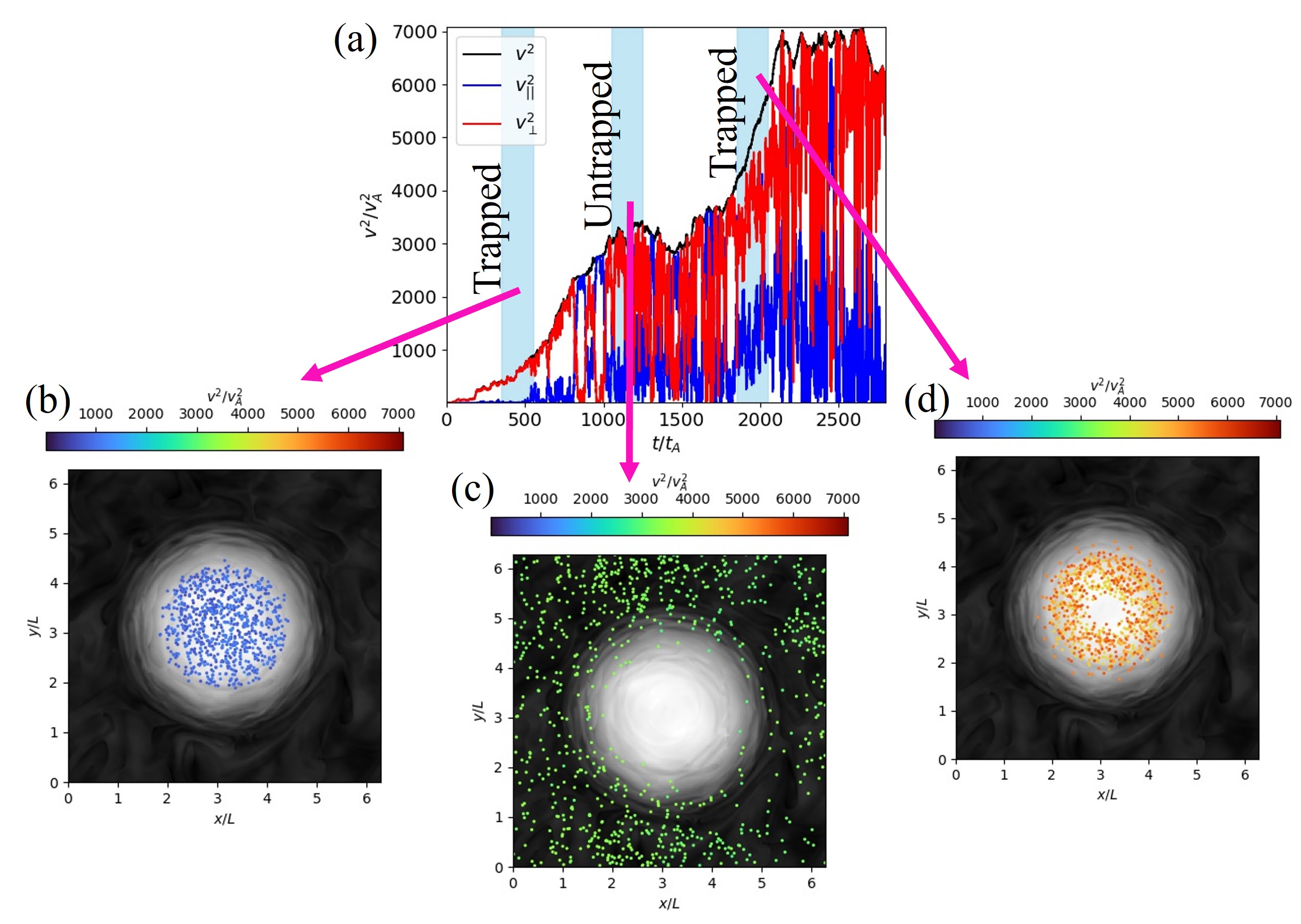}
\caption{(a): Total (black), perpendicular (red) and parallel (blue) energy for one of the most accelerated particles in the simulations. (b)-(d) Particle position during the shaded times in (a), coloured by particle energy. In the background, the magnetic field magnitude is shown in grey scale for the $z = L_{\rm box}/2$ plane.}
\label{fig:trapped_untrapped}
\end{figure*}

We further elucidate how particles are accelerated in the simulations by looking at their perpendicular and parallel energies. In the inset of Figure~\ref{fig:Enav}, it is shown that the mean perpendicular energy (dot-dashed line) is systematically larger than the parallel one (dashed line). In this analysis, performed for the case of uniform injection in the simulation domain, parallel and perpendicular velocities for each particle are computed with respect to the local magnetic field, which has been interpolated along the particle trajectory. Other injections are in qualitative agreement with what reported in the inset of Figure~\ref{fig:Enav}. At the beginning of the simulation, $\langle v_\perp^2 \rangle = 2 \langle v_\parallel^2\rangle$ as a result of the isotropic injection in velocity space. Suddenly and up to times $t\simeq 100 t_A$, the average perpendicular energy increases much faster than the parallel one as a result of fast acceleration processes related to particle trapping in the large-scale flux rope and leading to a preferentially perpendicular energization. For longer times, the average parallel energy increases as well for the underlying second-order process still energizing particles. Finally, for times $t>1000 t_A$ the isotropy of the velocity distribution is restored, i.e., $\langle v_\perp^2 \rangle \simeq 2 \langle v_\parallel^2\rangle$. We anticipate that trapping and fast perpendicular energization are recovered in all these different regimes we identified in the averaged energization process. However, at the beginning of the simulation, fast acceleration occurs in an environment weakly accelerated by slower processes, such as second-order energization, which eventually lead to the isotropization of the velocity distribution. Hence, the effect of trapping and consequent fast perpendicular energization is much more visible in the initial phase of the time history of the average kinetic energy. Such a result, in agreement with previous studies elucidating the role of particle trapping in turbulent structures in their subsequent acceleration~\citep[e.g.,][]{DmitrukEA04,dalena2014test, kowal2012particle, trotta2020fast, Li2021, pezzi2022relativistic}, has here been tested in a simulation dominated by a single, large-scale structure immersed in a turbulent background. Then, it becomes extremely interesting to address the interplay between particle trapping and energization in such a setting.

An important insight about the complexity of the energization mechanism is revealed in Figure~\ref{fig:par_perp}, where for the case of uniform injection we show the perpendicular and parallel energy history for the average sample (solid lines) and the same quantities for some individual particles (shaded lines) randomly selected from the entire sample. A rather complex picture, characterized by intense burst of acceleration occurring at different times for different particles, emerges. 

To clarify such a behavior, particles showing the most intense energization were selected, and their dynamics were studied in detail. Episodes of rapid acceleration have been found to be related to trapping in the turbulent flux rope. This is shown in Figure~\ref{fig:trapped_untrapped}, where the energy time-history of a single particle undergoing strong acceleration is shown together with its parallel and perpendicular energies (black, blue and red lines, respectively). It is interesting to note that this particle was trapped, then escaped the structure to be trapped later in another event of fast acceleration. This behavior, made possible here by the periodicity of the simulation, elucidates the possible behaviour for particles to undergo multiple trapping events at different structures. Such scenario, successfully invoked previously for multi-stage particle acceleration in localised networks of acceleration centers~\citep{Arzner&Vlahos2004, Vlahos2004,dalena2014test} will be investigated in future works involving more than one flux rope with different sizes. Rapid acceleration is driven by strong increases in perpendicular energy. Furthermore, trajectory analysis shows that such strong energization happens when trapping in the flux rope is active (Figure~\ref{fig:trapped_untrapped}(b)-(d)). Contrarily, when the particle is not trapped, no strong energisation is observed.

\section{Conclusions}
\label{sect:concl}

Turbulence in space plasmas is known to be structured and characterized by a cross-scale path of structures, which include magnetic islands, plasmoids, and turbulent eddies \citep{servidio2009magnetic}. In addition to the presence of these structures dynamically generated by turbulence, meso-scale structures such as flux ropes can also be of solar origin due to their association with coronal mass ejections \citep[see, e.g.,][]{hu2018automated,pecora2021parker,long2023eruption}. Although with different parameters, magnetic flux ropes are observed in the solar corona and are expected to be crucial for energizing particles \citep[see, e.g.,][]{antolin2010role, pinto2015soft, diazsuarez2021transition}. More generally, the presence of magnetic islands and plasmoids allows for fast particle energization as shown by several authors in rather different contexts \citep{drake2006electron,kowal2012particle,trotta2020fast,pezzi2022relativistic}.

There are a number of antecedents of the present work that introduce some of the concepts advanced in the present study; examples include: effects of trapping and escape on transport \citep{tooprakai2007temporary,tooprakai2016simulations}; particle exclusion \citep{seripienlert2010dropouts,pecora2021parker}; 
multistage acceleration \citep{dalena2014test}; and the importance of trapping in particle acceleration \citep{ambrosiano1988test}. In this work, we further elaborated these concepts by addressing the two following questions: (i) \textit{how does turbulence perturb --and is influenced by-- a large-scale flux rope?}; and (ii) \textit{how does a large-scale flux rope impact particle transport and energization?}. 
To this aim, by means of three-dimensional compressible MHD simulations performed with the \texttt{COHMPA} (``COmpressible Hall Magnetohydrodynamics simulations for Plasma Astrophysics'') algorithm, we built a twisted flux rope with the Grad-Shafranov technique and we perturbed it with large-scale fluctuations. Distinctive behaviors are observed in the cases of small or large amplitude of the initial perturbations. 
Indeed, as fluctuations are small, the turbulent transfer towards small scales is generally inhibited by the presence of the large-scale structure as evidenced by the scaling exponents $\zeta_q$ of the structure functions. 
On the other hand, in the case of intense perturbations, these mask the effect of the flux rope and scalings exponents reconcile with intermittent turbulence expectations. 
Similarly, the flux rope is of key importance in regulating particle transport and acceleration in the case in which it is weakly affected by turbulence. 
In this case, particles can be trapped by the large-scale structure that, hence, inhibits their transport in the directions perpendicular to the flux rope axis. 
Particles injected inside the flux rope have generally larger probability of being trapped within the structure. 
There, they can undergo episodes of intense and fast acceleration due to gradients in the electric field, mainly resulting from the presence of gradients in the plasma bulk speed.

This first study has focused on a flux rope in solar-wind conditions, though in this systems flux ropes exhibit a large variability of their fundamental parameters \citep{hu2018automated}. However, the above questions are rather general and other scenarios, characterized by different parameters, deserve a careful treatment. 
As an example, we here mention the interesting cases of coronal loops or flux ropes observed at different distances from the Sun. 
In the same spirit of understanding the interaction of Alfvenic wave packets \citep{Moffat69,Parker65,pezzi2017revisiting}, further studies will focus on the collisions of magnetic flux ropes, possibly orientated along different axes. 

To conclude, in future works we will explore a broader set of parameters by changing (i) the ratio of the turbulent correlation length with respect to the flux-rope size, (ii) the ratio of the flux-rope size and the particle gyroradius, and (iii) the typical scale of the large-scale flux-rope gradients. 
Moreover, we assumed that the turbulent flux rope achieves a stationary state in which we performed test-particle simulations. As discussed at length, it can be expected that flux ropes are long-lived structures which survive for several dynamical times. However, in future, we intend to perform simulations into non-static fields.

\begin{acknowledgements} 
This work has received funding from the European Unions Horizon2020 research and innovation programme under grant agreement No. 101004159 (\href{ www.serpentine-h2020.eu}{SERPENTINE}). 
L.S.-V. is supported by the Swedish Research Council (VR) Research Grant N. 2022-03352. 
W.H.M. is partially supported at the University of Delaware by US National Science Foundation grant AGS/PHYS-2108834 (NSFDOE) and by NASA Heliophysics GI grant (PSP) 80 NSSC21K1765. 
The simulations have been performed at the Newton cluster at University of Calabria and the work is supported by “Progetto STAR 2-PIR01 00008” (Italian Ministry of University and Research). 
SS acknowledges supercomputing resources and support from ICSC–Centro Nazionale di Ricerca in High Performance Computing, Big Data and Quantum Computing–and hosting entity, funded by European Union–NextGenerationEU.
\end{acknowledgements} 

\bibliographystyle{aa}
\bibliography{BIBLIO-living}

\begin{thebibliography}{102}
\expandafter\ifx\csname natexlab\endcsname\relax\def\natexlab#1{#1}\fi

\bibitem[{{Amato} \& {Casanova}(2021)}]{amato2021particle}
{Amato}, E. \& {Casanova}, S. 2021, Journal of Plasma Physics, 87, 845870101

\bibitem[{Ambrosiano {et~al.}(1988)Ambrosiano, Matthaeus, Goldstein, \&
  Plante}]{ambrosiano1988test}
Ambrosiano, J., Matthaeus, W.~H., Goldstein, M.~L., \& Plante, D. 1988, Journal
  of Geophysical Research, 93, 14\,383

\bibitem[{{Antolin} {et~al.}(2017){Antolin}, {De Moortel}, {Van Doorsselaere},
  \& {Yokoyama}}]{antolin2017observational}
{Antolin}, P., {De Moortel}, I., {Van Doorsselaere}, T., \& {Yokoyama}, T.
  2017, The Astrophysical Journal, 836, 219

\bibitem[{{Antolin} \& {Shibata}(2010)}]{antolin2010role}
{Antolin}, P. \& {Shibata}, K. 2010, The Astrophysical Journal, 712, 494

\bibitem[{{Antolin} {et~al.}(2014){Antolin}, {Yokoyama}, \& {Van
  Doorsselaere}}]{antolin2014fine}
{Antolin}, P., {Yokoyama}, T., \& {Van Doorsselaere}, T. 2014, The
  Astrophysical Journal Letters, 787, L22

\bibitem[{{Arzner} \& {Vlahos}(2004)}]{Arzner&Vlahos2004}
{Arzner}, K. \& {Vlahos}, L. 2004, \apjl, 605, L69

\bibitem[{{Benzi} {et~al.}(1993){Benzi}, {Ciliberto}, {Tripiccione}, {Baudet},
  {Massaioli}, \& {Succi}}]{benzi1993extended}
{Benzi}, R., {Ciliberto}, S., {Tripiccione}, R., {et~al.} 1993, \pre, 48, R29

\bibitem[{Birdsall \& Langdon(2004)}]{birdsall2004plasma}
Birdsall, C.~K. \& Langdon, A.~B. 2004, Plasma physics via computer simulation
  (CRC press)

\bibitem[{{Canuto} {et~al.}(2006){Canuto}, {Hussaini}, {Quarteroni}, \&
  {Zang}}]{canuto2006spectral}
{Canuto}, C., {Hussaini}, M.~Y., {Quarteroni}, A., \& {Zang}, T.~A. 2006,
  {Spectral Methods} (Springer-Verlag)

\bibitem[{Carbone \& Sorriso-Valvo(2014)}]{Carbone2014}
Carbone, F. \& Sorriso-Valvo, L. 2014, Eur. Phys. J. E, 37, 61

\bibitem[{Carbone(1993)}]{carbone1993}
Carbone, V. 1993, Phys. Rev. Lett., 71, 1546

\bibitem[{{Carbone} {et~al.}(2009){Carbone}, {Sorriso-Valvo}, \&
  {Marino}}]{carbone2009on}
{Carbone}, V., {Sorriso-Valvo}, L., \& {Marino}, R. 2009, EPL (Europhysics
  Letters), 88, 25001

\bibitem[{{Cristofari}(2021)}]{cristofari2021hunt}
{Cristofari}, P. 2021, Universe, 7, 324

\bibitem[{Dalena {et~al.}(2014)Dalena, Rappazzo, Dmitruk, Greco, \&
  Matthaeus}]{dalena2014test}
Dalena, S., Rappazzo, A.~F., Dmitruk, P., Greco, A., \& Matthaeus, W.~H. 2014,
  The Astrophysical Journal, 783, 143

\bibitem[{{D\'{\i}az-Su\'arez} \& {Soler}(2021)}]{diazsuarez2021transition}
{D\'{\i}az-Su\'arez}, S. \& {Soler}, R. 2021, Astronomy \& Astrophysics, 648,
  A22

\bibitem[{{D\'{\i}az-Su\'arez} \& {Soler}(2022)}]{diazsuarez2022transition}
{D\'{\i}az-Su\'arez}, S. \& {Soler}, R. 2022, Astronomy \& Astrophysics, 665,
  A113

\bibitem[{{Dmitruk} {et~al.}(2004){Dmitruk}, {Matthaeus}, \&
  {Seenu}}]{DmitrukEA04}
{Dmitruk}, P., {Matthaeus}, W.~H., \& {Seenu}, N. 2004, The Astrophysical
  Journal, 617, 667

\bibitem[{{Dmitruk} {et~al.}(2003){Dmitruk}, {Matthaeus}, {Seenu}, \&
  {Brown}}]{Dmitruk_al_2003}
{Dmitruk}, P., {Matthaeus}, W.~H., {Seenu}, N., \& {Brown}, M.~R. 2003,
  Astrophysical Journal Letters, 597, L81

\bibitem[{{Drake} {et~al.}(2006){Drake}, {Swisdak}, {Che}, \&
  {Shay}}]{drake2006electron}
{Drake}, J.~F., {Swisdak}, M., {Che}, H., \& {Shay}, M.~A. 2006, Nature, 443,
  553

\bibitem[{{Dresing} {et~al.}(2023){Dresing}, {Rodr{\'\i}guez-Garc{\'\i}a},
  {Jebaraj}, {Warmuth}, {Wallace}, {Balmaceda}, {Podladchikova}, {Strauss},
  {Kouloumvakos}, {Palmroos}, {Krupar}, {Gieseler}, {Xu}, {Mitchell}, {Cohen},
  {de Nolfo}, {Palmerio}, {Carcaboso}, {Kilpua}, {Trotta}, {Auster},
  {Asvestari}, {da Silva}, {Dr{\"o}ge}, {Getachew}, {G{\'o}mez-Herrero},
  {Grande}, {Heyner}, {Holmstr{\"o}m}, {Huovelin}, {Kartavykh}, {Laurenza},
  {Lee}, {Mason}, {Maksimovic}, {Mieth}, {Murakami}, {Oleynik}, {Pinto},
  {Pulupa}, {Richter}, {Rodr{\'\i}guez-Pacheco}, {S{\'a}nchez-Cano},
  {Schuller}, {Ueno}, {Vainio}, {Vecchio}, {Veronig}, \&
  {Wijsen}}]{dresing202317}
{Dresing}, N., {Rodr{\'\i}guez-Garc{\'\i}a}, L., {Jebaraj}, I.~C., {et~al.}
  2023, Astronomy \& Astrophysics, 674, A105

\bibitem[{{Dundovic} {et~al.}(2020){Dundovic}, {Pezzi}, {Blasi}, {Evoli}, \&
  {Matthaeus}}]{dundovic2020novel}
{Dundovic}, A., {Pezzi}, O., {Blasi}, P., {Evoli}, C., \& {Matthaeus}, W.~H.
  2020, Physical Review D, 102, 103016

\bibitem[{{Emonet} \& {Moreno-Insertis}(1998)}]{emonet1998physics}
{Emonet}, T. \& {Moreno-Insertis}, F. 1998, The Astrophysical Journal, 492, 804

\bibitem[{{Fermi}(1949)}]{Fermi1949}
{Fermi}, E. 1949, Physical Review, 75, 1169

\bibitem[{{Fermi}(1954)}]{Fermi1954}
{Fermi}, E. 1954, The Astrophysical Journal, 119, 1

\bibitem[{{Fisk} \& {Gloeckler}(2012)}]{fisk2012particle}
{Fisk}, L.~A. \& {Gloeckler}, G. 2012, Space Science Reviews, 173, 433

\bibitem[{Frigo(1999)}]{frigo1999fast}
Frigo, M. 1999, in Proceedings of the ACM SIGPLAN 1999 conference on
  Programming language design and implementation, 169--180

\bibitem[{Frisch(1995)}]{Frisch95}
Frisch, U. 1995, Turbulence: the legacy of AN Kolmogorov (Cambridge university
  press)

\bibitem[{{Gonz{\'a}lez} {et~al.}(2016){Gonz{\'a}lez}, {Dmitruk}, {Mininni}, \&
  {Matthaeus}}]{gonzalez2016compressibility}
{Gonz{\'a}lez}, C.~A., {Dmitruk}, P., {Mininni}, P.~D., \& {Matthaeus}, W.~H.
  2016, Physics of Plasmas, 23, 082305

\bibitem[{González {et~al.}(2017)González, Dmitruk, Mininni, \&
  Matthaeus}]{Gonzalez2017}
González, C.~A., Dmitruk, P., Mininni, P.~D., \& Matthaeus, W.~H. 2017, The
  Astrophysical Journal, 850, 19

\bibitem[{{Howson} {et~al.}(2017){Howson}, {De Moortel}, \&
  {Antolin}}]{howson2017energetics}
{Howson}, T.~A., {De Moortel}, I., \& {Antolin}, P. 2017, Astronomy \&
  Astrophysics, 607, A77

\bibitem[{Hu {et~al.}(2014)Hu, Qiu, Dasgupta, Khare, \&
  Webb}]{hu2014structures}
Hu, Q., Qiu, J., Dasgupta, B., Khare, A., \& Webb, G.~M. 2014, The
  Astrophysical Journal, 793, 53

\bibitem[{Hu {et~al.}(2018)Hu, Zheng, Chen, le~Roux, \& Zhao}]{hu2018automated}
Hu, Q., Zheng, J., Chen, Y., le~Roux, J., \& Zhao, L. 2018, The Astrophysical
  Journal Supplement Series, 239, 12

\bibitem[{{Karampelas} {et~al.}(2017){Karampelas}, {Van Doorsselaere}, \&
  {Antolin}}]{karampelas2017heating}
{Karampelas}, K., {Van Doorsselaere}, T., \& {Antolin}, P. 2017, Astronomy \&
  Astrophysics, 604, A130

\bibitem[{{Khabarova} {et~al.}(2021){Khabarova}, {Malandraki}, {Malova},
  {Kislov}, {Greco}, {Bruno}, {Pezzi}, {Servidio}, {Li}, {Matthaeus}, {Le
  Roux}, {Engelbrecht}, {Pecora}, {Zelenyi}, {Obridko}, \&
  {Kuznetsov}}]{khabarova2021currentsheets}
{Khabarova}, O., {Malandraki}, O., {Malova}, H., {et~al.} 2021, Space Science
  Reviews, 217, 38

\bibitem[{{Khabarova} {et~al.}(2015){Khabarova}, {Zank}, {Li}, {le Roux},
  {Webb}, {Dosch}, \& {Malandraki}}]{khabarova2015smallscale}
{Khabarova}, O., {Zank}, G.~P., {Li}, G., {et~al.} 2015, The Astrophysical
  Journal, 808, 181

\bibitem[{{Khabarova} {et~al.}(2016){Khabarova}, {Zank}, {Li}, {Malandraki},
  {le Roux}, \& {Webb}}]{khabarova2016smallscale}
{Khabarova}, O.~V., {Zank}, G.~P., {Li}, G., {et~al.} 2016, The Astrophysical
  Journal, 827, 122

\bibitem[{{Kilpua} {et~al.}(2023){Kilpua}, {Vainio}, {Cohen}, {Dresing},
  {Good}, {Ruohotie}, {Trotta}, {Bale}, {Christian}, {Hill}, {McComas},
  {McNutt}, \& {Schwadron}}]{Kilpua2023sheath}
{Kilpua}, E., {Vainio}, R., {Cohen}, C., {et~al.} 2023, \apss, 368, 66

\bibitem[{{Kolmogorov}(1941)}]{Kolmogorov41a}
{Kolmogorov}, A. 1941, Akademiia Nauk SSSR Doklady, 30, 301

\bibitem[{{Kowal} {et~al.}(2011){Kowal}, {de Gouveia Dal Pino}, \&
  {Lazarian}}]{kowal2011magnetohydrodynamic}
{Kowal}, G., {de Gouveia Dal Pino}, E.~M., \& {Lazarian}, A. 2011, The
  Astrophysical Journal, 735, 102

\bibitem[{{Kowal} {et~al.}(2012){Kowal}, {de Gouveia Dal Pino}, \&
  {Lazarian}}]{kowal2012particle}
{Kowal}, G., {de Gouveia Dal Pino}, E.~M., \& {Lazarian}, A. 2012, Physical
  Review Letters, 108, 241102

\bibitem[{{Kraichnan}(1965)}]{Kraichnan65}
{Kraichnan}, R. 1965, Physics of Fluids, 8, 1385

\bibitem[{{Krittinatham} \& {Ruffolo}(2009)}]{krittinatham2009drift}
{Krittinatham}, W. \& {Ruffolo}, D. 2009, The Astrophysical Journal, 704, 831

\bibitem[{{Lazarian} {et~al.}(2020){Lazarian}, {Eyink}, {Jafari}, {Kowal},
  {Li}, {Xu}, \& {Vishniac}}]{lazarian20203D}
{Lazarian}, A., {Eyink}, G.~L., {Jafari}, A., {et~al.} 2020, Physics of
  Plasmas, 27, 012305

\bibitem[{{Lazarian} {et~al.}(2012){Lazarian}, {Vlahos}, {Kowal}, {Yan},
  {Beresnyak}, \& {de Gouveia Dal Pino}}]{lazarian2012turbulence}
{Lazarian}, A., {Vlahos}, L., {Kowal}, G., {et~al.} 2012, \ssr, 173, 557

\bibitem[{{le Roux} {et~al.}(2019){le Roux}, Webb, Khabarova, Zhao, \&
  Adhikari}]{leroux2019modeling}
{le Roux}, J.~A., Webb, G.~M., Khabarova, O.~V., Zhao, L.-L., \& Adhikari, L.
  2019, The Astrophysical Journal, 887, 77

\bibitem[{{le Roux} {et~al.}(2018){le Roux}, {Zank}, \&
  {Khabarova}}]{leroux2018selfconsistent}
{le Roux}, J.~A., {Zank}, G.~P., \& {Khabarova}, O.~V. 2018, The Astrophysical
  Journal, 864, 158

\bibitem[{{le Roux} {et~al.}(2015){le Roux}, {Zank}, {Webb}, \&
  {Khabarova}}]{leroux2015kinetic}
{le Roux}, J.~A., {Zank}, G.~P., {Webb}, G.~M., \& {Khabarova}, O. 2015, The
  Astrophysical Journal, 801, 112

\bibitem[{{Lemoine}(2022)}]{lemoine2022first}
{Lemoine}, M. 2022, \prl, 129, 215101

\bibitem[{Li {et~al.}(2021)Li, Guo, \& Liu}]{Li2021}
Li, X., Guo, F., \& Liu, Y.-H. 2021, Physics of Plasmas, 28, 052905

\bibitem[{{Long} {et~al.}(2023){Long}, {Green}, {Pecora}, {Brooks}, {Strecker},
  {Orozco-Su{\'a}rez}, {Hayes}, {Davies}, {Amerstorfer}, {Mierla}, {Lario},
  {Berghmans}, {Zhukov}, \& {R{\"u}disser}}]{long2023eruption}
{Long}, D.~M., {Green}, L.~M., {Pecora}, F., {et~al.} 2023, The Astrophysical
  Journal, 955, 152

\bibitem[{{Magyar} {et~al.}(2015){Magyar}, {Van Doorsselaere}, \&
  {Marcu}}]{magyar2015numerical}
{Magyar}, N., {Van Doorsselaere}, T., \& {Marcu}, A. 2015, Astronomy and
  Astrophysics, 582, A117

\bibitem[{{Maiorano} {et~al.}(2020){Maiorano}, {Settino}, {Malara}, {Pezzi},
  {Pucci}, \& {Valentini}}]{maiorano2020kinetic}
{Maiorano}, T., {Settino}, A., {Malara}, F., {et~al.} 2020, Journal of Plasma
  Physics, 86, 825860202

\bibitem[{{Malandraki} {et~al.}(2019){Malandraki}, {Khabarova}, {Bruno},
  {Zank}, {Li}, {Jackson}, {Bisi}, {Greco}, {Pezzi}, {Matthaeus}, {Chasapis
  Giannakopoulos}, {Servidio}, {Malova}, {Kislov}, {Effenberger}, {le Roux},
  {Chen}, {Hu}, \& {Engelbrecht}}]{malandraki2019current}
{Malandraki}, O., {Khabarova}, O., {Bruno}, R., {et~al.} 2019, The
  Astrophysical Journal, 881, 116

\bibitem[{{Malara} {et~al.}(2023){Malara}, {Perri}, {Giacalone}, \&
  {Zimbardo}}]{malara2023energetic}
{Malara}, F., {Perri}, S., {Giacalone}, J., \& {Zimbardo}, G. 2023, Astronomy
  and Astrophysics, 677, A69

\bibitem[{{Malara} {et~al.}(2021){Malara}, {Perri}, \&
  {Zimbardo}}]{malara2021charged}
{Malara}, F., {Perri}, S., \& {Zimbardo}, G. 2021, Physical Review E, 104,
  025208

\bibitem[{Marino \& Sorriso-Valvo(2023)}]{marino2023scaling}
Marino, R. \& Sorriso-Valvo, L. 2023, Physics Reports, 1006, 1, scaling laws
  for the energy transfer in space plasma turbulence

\bibitem[{{Matthaeus} {et~al.}(2015){Matthaeus}, {Wan}, {Servidio}, {Greco},
  {Osman}, {Oughton}, \& {Dmitruk}}]{matthaeus2015intermittency}
{Matthaeus}, W.~H., {Wan}, M., {Servidio}, S., {et~al.} 2015, Philosophical
  Transactions of the Royal Society of London Series A, 373, 20140154

\bibitem[{{McComas} {et~al.}(2023){McComas}, {Sharma}, {Christian}, {Cohen},
  {Desai}, {Hill}, {Khoo}, {Matthaeus}, {Mitchell}, {Pecora}, {Rankin},
  {Schwadron}, {Szalay}, {Shen}, {Braga}, {Mostafavi}, \&
  {Bale}}]{mccomas2023parker}
{McComas}, D.~J., {Sharma}, T., {Christian}, E.~R., {et~al.} 2023, The
  Astrophysical Journal, 943, 71

\bibitem[{Meneveau \& Sreenivasan(1987)}]{Meneveau87}
Meneveau, C. \& Sreenivasan, K. 1987, Physical review letters, 59, 1424

\bibitem[{Moffatt(1969)}]{Moffat69}
Moffatt, H.~K. 1969, Journal of Fluid Mechanics, 35, 117

\bibitem[{Nakanotani {et~al.}(2021)Nakanotani, Zank, \& Zhao}]{Nakanotani2021}
Nakanotani, M., Zank, G.~P., \& Zhao, L.-L. 2021, The Astrophysical Journal,
  922, 219

\bibitem[{{Oka} {et~al.}(2010){Oka}, {Phan}, {Krucker}, {Fujimoto}, \&
  {Shinohara}}]{oka2010electron}
{Oka}, M., {Phan}, T.~D., {Krucker}, S., {Fujimoto}, M., \& {Shinohara}, I.
  2010, The Astrophysical Journal, 714, 915

\bibitem[{Parker(1965)}]{Parker65}
Parker, E.~N. 1965, Space Sci. Rev., 4, 666

\bibitem[{Pecora {et~al.}(2018)Pecora, Servidio, Greco, Matthaeus, Burgess,
  Haynes, Carbone, \& Veltri}]{pecora2017ion}
Pecora, F., Servidio, S., Greco, A., {et~al.} 2018, Journal of Plasma Physics,
  84, 725840601

\bibitem[{{Pecora} {et~al.}(2021){Pecora}, {Servidio}, {Greco}, {Matthaeus},
  {McComas}, {Giacalone}, {Joyce}, {Getachew}, {Cohen}, {Leske}, {Wiedenbeck},
  {McNutt}, {Hill}, {Mitchell}, {Christian}, {Roelof}, {Schwadron}, \&
  {Bale}}]{pecora2021parker}
{Pecora}, F., {Servidio}, S., {Greco}, A., {et~al.} 2021, Monthly Notices of
  the Royal Astronomical Society, 508, 2114

\bibitem[{Perri {et~al.}(2017)Perri, Servidio, Vaivads, \&
  Valentini}]{perri2017numerical}
Perri, S., Servidio, S., Vaivads, A., \& Valentini, F. 2017, The Astrophysical
  Journal Supplement Series, 231, 4

\bibitem[{{Pezzi} {et~al.}(2022){Pezzi}, {Blasi}, \&
  {Matthaeus}}]{pezzi2022relativistic}
{Pezzi}, O., {Blasi}, P., \& {Matthaeus}, W.~H. 2022, \apj, 928, 25

\bibitem[{Pezzi {et~al.}(2017)Pezzi, Parashar, Servidio, Valentini, V\'asconez,
  Yang, Malara, Matthaeus, \& Veltri}]{pezzi2017revisiting}
Pezzi, O., Parashar, T.~N., Servidio, S., {et~al.} 2017, The Astrophysical
  Journal, 834, 166

\bibitem[{{Pezzi} {et~al.}(2021){Pezzi}, {Pecora}, {Le Roux}, {Engelbrecht},
  {Greco}, {Servidio}, {Malova}, {Khabarova}, {Malandraki}, {Bruno},
  {Matthaeus}, {Li}, {Zelenyi}, {Kislov}, {Obridko}, \&
  {Kuznetsov}}]{pezzi2021currentsheets}
{Pezzi}, O., {Pecora}, F., {Le Roux}, J., {et~al.} 2021, Space Science Reviews,
  217, 39

\bibitem[{{Pinto} {et~al.}(2016){Pinto}, {Gordovskyy}, {Browning}, \&
  {Vilmer}}]{pinto2016thermal}
{Pinto}, R.~F., {Gordovskyy}, M., {Browning}, P.~K., \& {Vilmer}, N. 2016,
  Astronomy \& Astrophysics, 585, A159

\bibitem[{{Pinto} {et~al.}(2015){Pinto}, {Vilmer}, \& {Brun}}]{pinto2015soft}
{Pinto}, R.~F., {Vilmer}, N., \& {Brun}, A.~S. 2015, Astronomy \& Astrophysics,
  576, A37

\bibitem[{{Pisokas} {et~al.}(2018){Pisokas}, {Vlahos}, \&
  {Isliker}}]{pisokas2018synergy}
{Pisokas}, T., {Vlahos}, L., \& {Isliker}, H. 2018, The Astrophysical Journal,
  852, 64

\bibitem[{{Politano} \& {Pouquet}(1998)}]{politano1998vonkarman}
{Politano}, H. \& {Pouquet}, A. 1998, Physical Review E, 57, R21

\bibitem[{Quijia {et~al.}(2021)Quijia, Fraternale, Stawarz, Vásconez, Perri,
  Marino, Yordanova, \& Sorriso-Valvo}]{Quijia2021}
Quijia, P., Fraternale, F., Stawarz, J.~E., {et~al.} 2021, Monthly Notices of
  the Royal Astronomical Society, 503, 4815

\bibitem[{Reames(1999)}]{reames99particle}
Reames, D.~V. 1999, Space Science Reviews, 90, 413

\bibitem[{{Retin{\`o}} {et~al.}(2022){Retin{\`o}}, {Khotyaintsev}, {Le Contel},
  {Marcucci}, {Plaschke}, {Vaivads}, {Angelopoulos}, {Blasi}, {Burch}, {De
  Keyser}, {Dunlop}, {Dai}, {Eastwood}, {Fu}, {Haaland}, {Hoshino},
  {Johlander}, {Kepko}, {Kucharek}, {Lapenta}, {Lavraud}, {Malandraki},
  {Matthaeus}, {McWilliams}, {Petrukovich}, {Pin{\c{c}}on}, {Saito},
  {Sorriso-Valvo}, {Vainio}, \& {Wimmer-Schweingruber}}]{retino2022particle}
{Retin{\`o}}, A., {Khotyaintsev}, Y., {Le Contel}, O., {et~al.} 2022,
  Experimental Astronomy, 54, 427

\bibitem[{{R{\'e}ville} {et~al.}(2022){R{\'e}ville}, {Fargette}, {Rouillard},
  {Lavraud}, {Velli}, {Strugarek}, {Parenti}, {Brun}, {Shi}, {Kouloumvakos},
  {Poirier}, {Pinto}, {Louarn}, {Fedorov}, {Owen}, {G{\'e}not}, {Horbury},
  {Laker}, {O'Brien}, {Angelini}, {Fauchon-Jones}, \&
  {Kasper}}]{reville2022flux}
{R{\'e}ville}, V., {Fargette}, N., {Rouillard}, A.~P., {et~al.} 2022, Astronomy
  \& Astrophysics, 659, A110

\bibitem[{{Ripperda} {et~al.}(2018){Ripperda}, {Bacchini}, {Teunissen}, {Xia},
  {Porth}, {Sironi}, {Lapenta}, \& {Keppens}}]{ripperda2018comprehensive}
{Ripperda}, B., {Bacchini}, F., {Teunissen}, J., {et~al.} 2018, The
  Astrophysical Journal Supplement Series, 235, 21

\bibitem[{{Seripienlert} {et~al.}(2010){Seripienlert}, {Ruffolo}, {Matthaeus},
  \& {Chuychai}}]{seripienlert2010dropouts}
{Seripienlert}, A., {Ruffolo}, D., {Matthaeus}, W.~H., \& {Chuychai}, P. 2010,
  The Astrophysical Journal, 711, 980

\bibitem[{Servidio {et~al.}(2011)Servidio, Dmitruk, Greco, Wan, Donato, Cassak,
  Shay, Carbone, \& Matthaeus}]{servidio2011magnetic}
Servidio, S., Dmitruk, P., Greco, A., {et~al.} 2011, Nonlinear Processes in
  Geophysics, 18, 675

\bibitem[{{Servidio} {et~al.}(2016){Servidio}, {Haynes}, {Matthaeus},
  {Burgess}, {Carbone}, \& {Veltri}}]{servidio2016explosive}
{Servidio}, S., {Haynes}, C.~T., {Matthaeus}, W.~H., {et~al.} 2016, Physical
  Review Letters, 117, 095101

\bibitem[{Servidio {et~al.}(2008)Servidio, Matthaeus, \&
  Dmitruk}]{servidio2008depression}
Servidio, S., Matthaeus, W.~H., \& Dmitruk, P. 2008, Phys. Rev. Lett., 100,
  095005

\bibitem[{{Servidio} {et~al.}(2009){Servidio}, {Matthaeus}, {Shay}, {Cassak},
  \& {Dmitruk}}]{servidio2009magnetic}
{Servidio}, S., {Matthaeus}, W.~H., {Shay}, M.~A., {Cassak}, P.~A., \&
  {Dmitruk}, P. 2009, Physical Review Letters, 102, 115003

\bibitem[{Shebalin {et~al.}(1983)Shebalin, Matthaeus, \&
  Montgomery}]{Shebalin83}
Shebalin, J.~V., Matthaeus, W.~H., \& Montgomery, D. 1983, J. Plasma Phys., 29,
  525

\bibitem[{{Sorriso-Valvo} {et~al.}(2007){Sorriso-Valvo}, {Marino}, {Carbone},
  {Noullez}, {Lepreti}, {Veltri}, {Bruno}, {Bavassano}, \&
  {Pietropaolo}}]{sorriso2007observation}
{Sorriso-Valvo}, L., {Marino}, R., {Carbone}, V., {et~al.} 2007, Physical
  Review Letters, 99, 115001

\bibitem[{{Sorriso-Valvo} {et~al.}(2018){Sorriso-Valvo}, {Perrone}, {Pezzi},
  {Valentini}, {Servidio}, {Zouganelis}, \& {Veltri}}]{sorriso2018local}
{Sorriso-Valvo}, L., {Perrone}, D., {Pezzi}, O., {et~al.} 2018, Journal of
  Plasma Physics, 84, 725840201

\bibitem[{{Srivastava} {et~al.}(2010){Srivastava}, {Zaqarashvili}, {Kumar}, \&
  {Khodachenko}}]{srivastava2010observation}
{Srivastava}, A.~K., {Zaqarashvili}, T.~V., {Kumar}, P., \& {Khodachenko},
  M.~L. 2010, The Astrophysical Journal, 715, 292

\bibitem[{{Terradas} {et~al.}(2008){Terradas}, {Andries}, {Goossens},
  {Arregui}, {Oliver}, \& {Ballester}}]{terradas2008nonlinear}
{Terradas}, J., {Andries}, J., {Goossens}, M., {et~al.} 2008, The Astrophysical
  Journal Letters, 687, L115

\bibitem[{Tooprakai {et~al.}(2007)Tooprakai, Chuychai, Minnie, Ruffolo, Bieber,
  \& Matthaeus}]{tooprakai2007temporary}
Tooprakai, P., Chuychai, P., Minnie, J., {et~al.} 2007, Geophysical Research
  Letters, 34

\bibitem[{{Tooprakai} {et~al.}(2016){Tooprakai}, {Seripienlert}, {Ruffolo},
  {Chuychai}, \& {Matthaeus}}]{tooprakai2016simulations}
{Tooprakai}, P., {Seripienlert}, A., {Ruffolo}, D., {Chuychai}, P., \&
  {Matthaeus}, W.~H. 2016, The Astrophysical Journal, 831, 195

\bibitem[{{Trotta} \& {Burgess}(2019)}]{Trotta2019}
{Trotta}, D. \& {Burgess}, D. 2019, \mnras, 482, 1154

\bibitem[{{Trotta} {et~al.}(2020){Trotta}, {Franci}, {Burgess}, \&
  {Hellinger}}]{trotta2020fast}
{Trotta}, D., {Franci}, L., {Burgess}, D., \& {Hellinger}, P. 2020, The
  Astrophysical Journal, 894, 136

\bibitem[{Trotta {et~al.}(2022)Trotta, Pecora, Settino, Perrone, Hietala,
  Horbury, Matthaeus, Burgess, Servidio, \& Valentini}]{Trotta2022a}
Trotta, D., Pecora, F., Settino, A., {et~al.} 2022, The Astrophysical Journal,
  933, 167

\bibitem[{{Valentini, F.} {et~al.}(2017){Valentini, F.}, {V\'asconez, C. L.},
  {Pezzi, O.}, {Servidio, S.}, {Malara, F.}, \& {Pucci,
  F.}}]{valentini2017transition}
{Valentini, F.}, {V\'asconez, C. L.}, {Pezzi, O.}, {et~al.} 2017, "Astronomy
  and Astrophysics", 599, "A8"

\bibitem[{V\'asconez {et~al.}(2015)V\'asconez, Pucci, Valentini, Servidio,
  Matthaeus, \& Malara}]{vasconez2015kinetic}
V\'asconez, C.~L., Pucci, F., Valentini, F., {et~al.} 2015, The Astrophysical
  Journal, 815, 7

\bibitem[{{Viall} {et~al.}(2021){Viall}, {DeForest}, \&
  {Kepko}}]{viall2021mesoscale}
{Viall}, N.~M., {DeForest}, C.~E., \& {Kepko}, L. 2021, Frontiers in Astronomy
  and Space Sciences, 8, 139

\bibitem[{Vlahos {et~al.}(2004)Vlahos, Isliker, \& Lepreti}]{Vlahos2004}
Vlahos, L., Isliker, H., \& Lepreti, F. 2004, The Astrophysical Journal, 608,
  540

\bibitem[{{Wan} {et~al.}(2009){Wan}, {Servidio}, {Oughton}, \&
  {Matthaeus}}]{wan2009thirdorder}
{Wan}, M., {Servidio}, S., {Oughton}, S., \& {Matthaeus}, W.~H. 2009, Physics
  of Plasmas, 16, 090703

\bibitem[{{Wijsen} {et~al.}(2022){Wijsen}, {Aran}, {Scolini}, {Lario},
  {Afanasiev}, {Vainio}, {Sanahuja}, {Pomoell}, \&
  {Poedts}}]{wijsen2022observation}
{Wijsen}, N., {Aran}, A., {Scolini}, C., {et~al.} 2022, Astronomy and
  Astrophysics, 659, A187

\bibitem[{Yordanova {et~al.}(2008)Yordanova, Vaivads, Andr{\'e}, Buchert, \&
  V{\"o}r{\"o}s}]{Yordanova08}
Yordanova, E., Vaivads, A., Andr{\'e}, M., Buchert, S., \& V{\"o}r{\"o}s, Z.
  2008, Physical Review Letters, 100, 205003

\bibitem[{Zank {et~al.}(2015)Zank, Hunana, Mostafavi, le~Roux, Li, Webb, \&
  Khabarova}]{Zank2015}
Zank, G.~P., Hunana, P., Mostafavi, P., {et~al.} 2015, Journal of Physics:
  Conference Series, 642, 012031

\bibitem[{Zank {et~al.}(2021)Zank, Nakanotani, Zhao, Du, Adhikari, Che, \&
  le~Roux}]{Zank2021}
Zank, G.~P., Nakanotani, M., Zhao, L.~L., {et~al.} 2021, The Astrophysical
  Journal, 913, 127

\end{thebibliography}

\begin{appendix}

\section{Grad-Shafranov equilibria}
\label{App:GS}

\begin{figure}[!htb]
\centering
\begin{minipage}{0.33\textwidth}
\includegraphics[width=\textwidth]{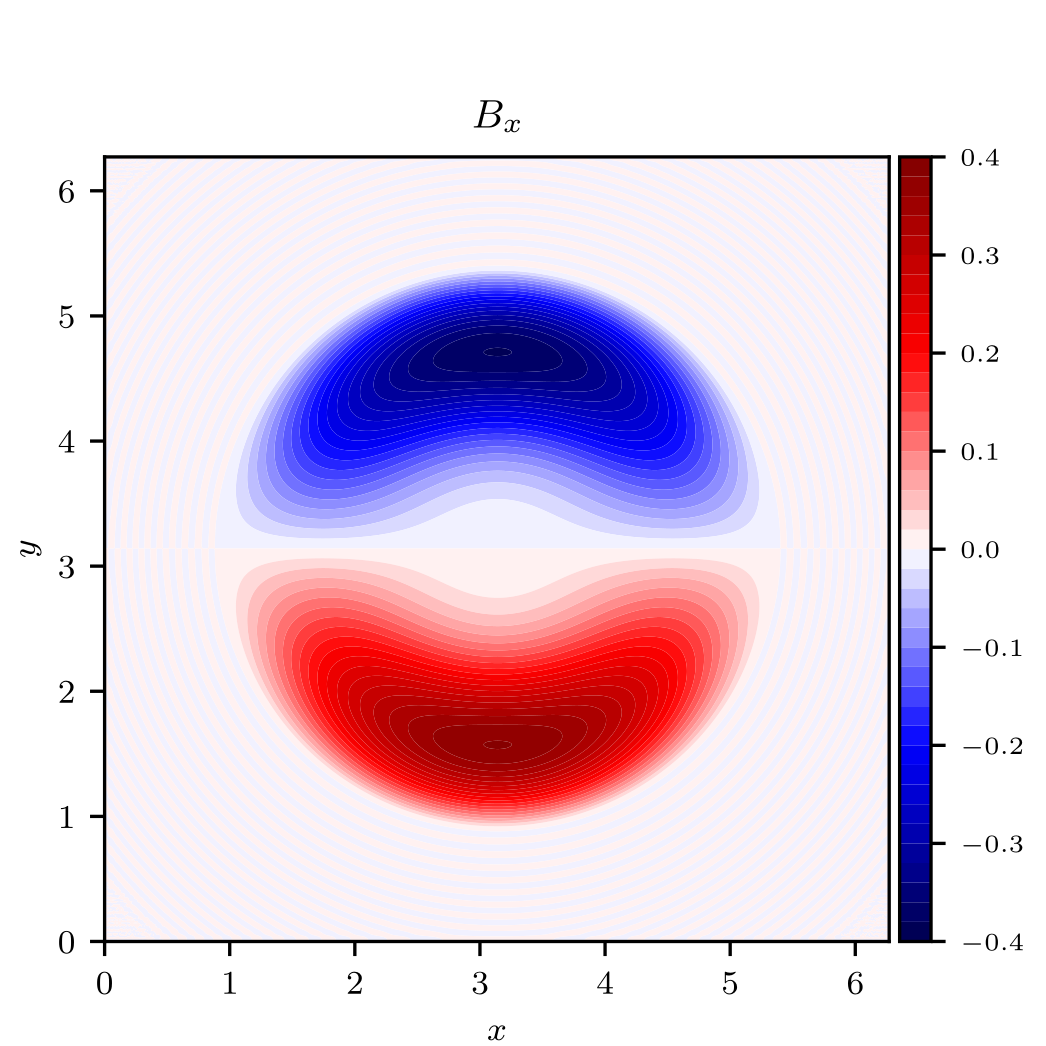}
\end{minipage}
\begin{minipage}{0.33\textwidth}
\includegraphics[width=\textwidth]{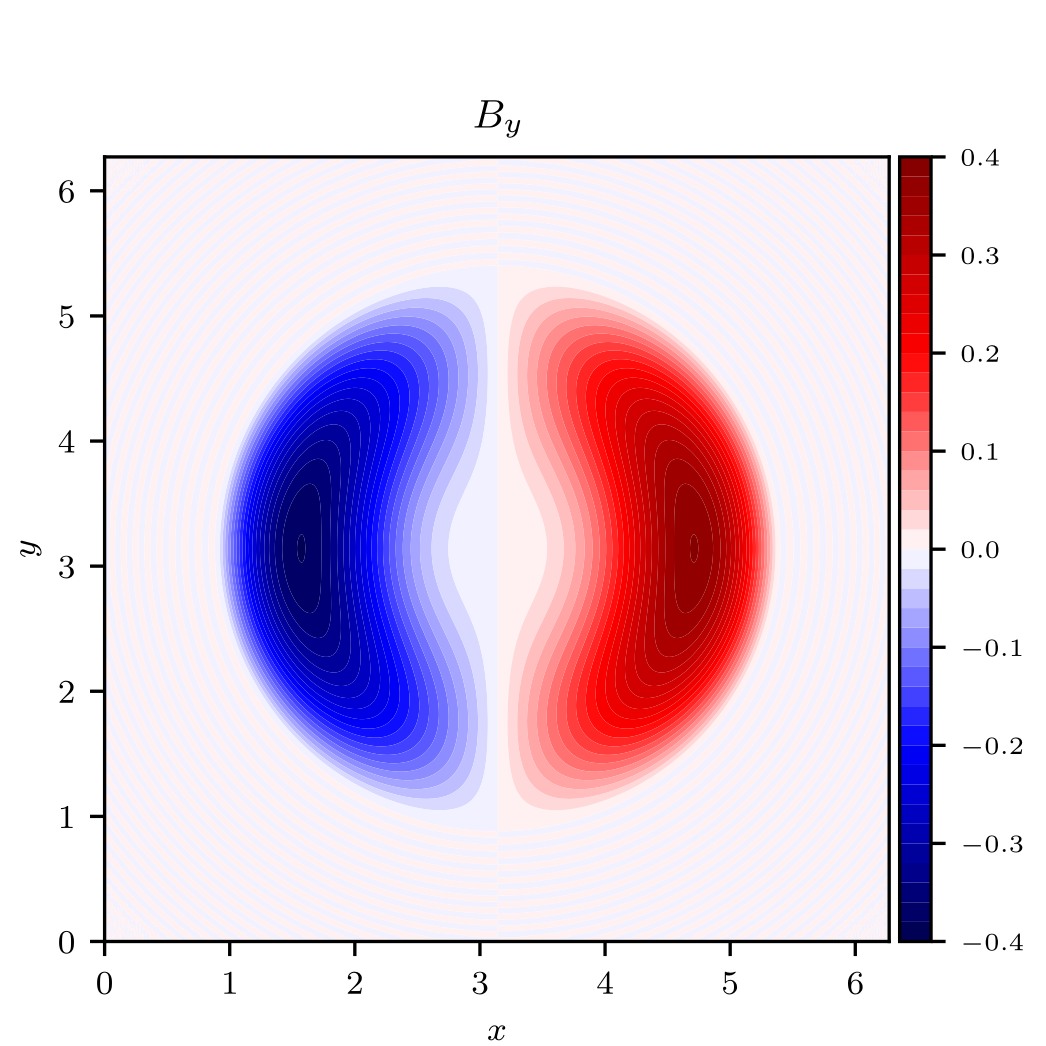}
\end{minipage}
\begin{minipage}{0.33\textwidth}
\includegraphics[width=\textwidth]{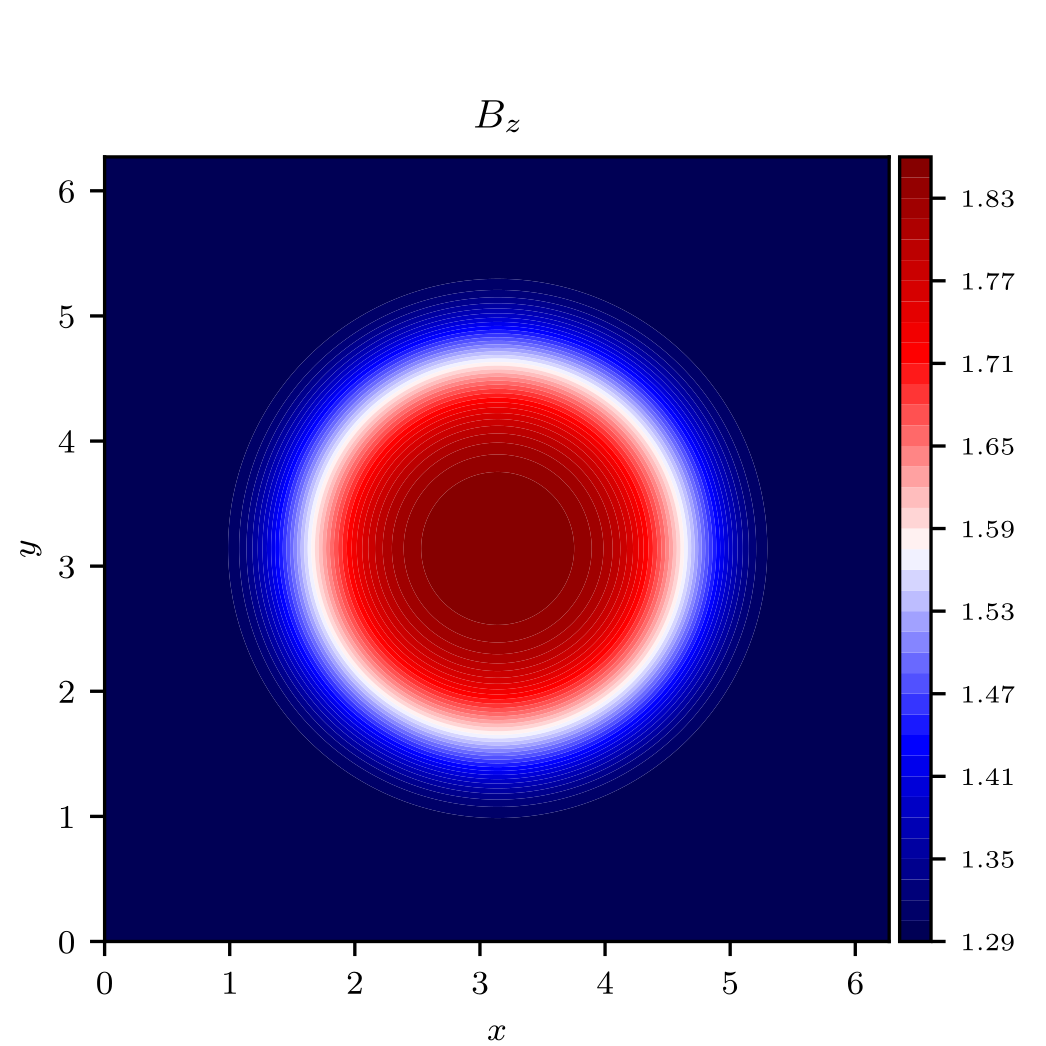}
\end{minipage}
\caption{Contour plots of the quilibrium magnetic field. From top to bottom: $B_x$, $B_y$, and $B_z$.}
\label{fig:EQ_2DmapsB}
\end{figure}

The Grad--Shafranov equation is an equilibrium equation og ideal MHD capable of describing several configurations, including the reversed field pinch in fusion devices or the solar prominences. Assuming a static plasma in the absence of visco-resistive forces, the momentum equation (normalized as in Sect. \ref{sect:nummod}) reads as:
\begin{equation}
 {\bm j}\times{\bm B} = {\bm \nabla} P,
 \label{eq:GS}
\end{equation}
where $\bm j={\bm \nabla}\times{\bm B}$ is the current density and $P$ is the kinetic pressure. Equation \ref{eq:GS} is solved by decomposing the magnetic field via the Euler potentials and assuming a relation between the poloidal and the toroidal fluxes:
\begin{equation}
{\bm B} =A \psi {\bm {\hat z}} + {\bm \nabla \psi}\times{\bm {\hat z}}, 
 \label{eq:dec}
\end{equation}
The toroidal flux is hence proportional to the stream function $\psi$ (being $A$ constant, for simplicity), which in turn is poloidal flux. 

Here, we focus on the 2.5D configuration, $A \psi\equiv A \psi(x, y)$ is the out-of-plane magnetic vector potential $a_z$, being ${\bm \nabla}\equiv (\partial_x, \partial_y)$. By substituting Eq.(\ref{eq:dec}) in (\ref{eq:GS}), one gets the following PDEs:
\begin{eqnarray}
-\partial_x \psi \left[\partial^2_{xx}\psi + \partial^2_{yy}\psi + A^2 \psi  \right]=\partial_x P, \label{eq:mess1} \\
-\partial_y \psi \left[\partial^2_{xx}\psi + \partial^2_{yy}\psi + A^2 \psi  \right]=\partial_y P.
\label{eq:mess2}
\end{eqnarray}
The term in the square parenthesis is essentially the Helmholtz equation, which is null in the case of force-free states. The main difference is that the GS equation keeps the parallel component of the magnetic field. Hence, magnetic field lines characterized by helicoidal states are allowed. This new degree of freedom provides a richer variety of solutions.

Equations (\ref{eq:mess1}--\ref{eq:mess2}) are solved for a given differentiable stream function $\psi$ to provide the kinetic pressure $P$. Indeed, these equations correspond a Poisson equation for $P$:
\begin{equation}
\nabla^2 P = G,  
\label{eq:PoisP}
\end{equation}
where 
\begin{equation}
    G= \partial_x{\left\{  -\partial_x \psi \left[\partial^2_{xx}\psi + \partial^2_{yy}\psi + A^2 \psi  \right] \right\}} +
\partial_y{\left\{ -\partial_y \psi \left[\partial^2_{xx}\psi + \partial^2_{yy}\psi + A^2 \psi  \right] \right\}}
\end{equation}

%``https://en.wikipedia.org/wiki/Window$-$function''
With the purpose of localizing the magnetic flux in a particular region of the computational domain, we build the flux $\psi$ through the Blackman-Nuttall window. In particular, we initially considered the following 1D axisimmetric profile:
\begin{equation}
\psi( x,y ) = \psi_0 +\psi_1 \cos( r - \pi ) + \psi_2 \cos( 2 r - \pi ) +\psi_3 \cos ( 3 r - \pi) \, \, ,     
\end{equation}
being $\psi_0=0.355768$, $\psi_1=-0.487396$, $\psi_2=0.144232$, and $\psi_3=-0.012604$ opportune coefficients which control the amplitude and the shear layer width of the flux function, while $r=1.4\sqrt{(x-x_0)^2+(y-y_0)^2}$ with $x_0=y_0=\pi$. Then, we renormalized $\psi$ such that $\langle\psi\rangle=1.0$, while its amplitude is $0.4$. Moreover, $A=1.4$. Within these parameters, the resulting magnetic vector potential is the one reported in Fig. \ref{fig:EQ_2Dmaps} (right). The magnetic field associated with such a potential evaluated through Eq. (\ref{eq:dec}) is shown in Fig. \ref{fig:EQ_2DmapsB}. The in-plane magnetic field components have a dipolar structure, while the (toroidal-like) field $B_z$ is much more intense. Given the stream function $\psi$, Eq. (\ref{eq:PoisP}) is solved by exploiting the periodic boundary conditions and setting the minimum pressure to the value $P_0=0.5$. Finally, by using the adiabatic closure for an ideal gas, we can get the density $\rho$ and the pressure $T$.

The choice of $\psi, A$, $P_0$ (as well as $\rho_0$ and $T_0$) is rather arbitrary. We selected the above parameters for achieving a configuration qualitatively similar to in-situ observations in the solar wind. Different configurations will be considered in a separate work.

\end{appendix}

\end{document}